\documentclass[longbibliography]{revtex4-2}

\usepackage{graphicx}
\usepackage{dcolumn}
\usepackage{bm}
\usepackage{siunitx} 
\usepackage{hyperref}
\usepackage{graphicx}
\usepackage{epstopdf}
\usepackage{rotating}
\usepackage{array}
\usepackage{color}
\usepackage{slashed}
\usepackage{multirow}
\usepackage{amsmath}
\usepackage[makeroom]{cancel}
\usepackage{tabularx}
\usepackage{cancel}

\begin{document}

\title{\Large The  $Z_3$ soft breaking in the I(2+1)HDM\\ and its cosmological probes }

\author{M. A. Arroyo-Ure\~na}
\email[]{marco.arroyo@fcfm.buap.mx}
\affiliation{Facultad de Ciencias F\'isico-Matem\'aticas, Benem\'erita Universidad Aut\'onoma de Puebla, C.P. 72570, Puebla, M\'exico}

\author{J. Hern\'andez-S\'anchez}
\email[]{jaime.hernandez@correo.buap.mx}
\affiliation{Facultad de Ciencias de la Electr\'onica, Benem\'erita Universidad Aut\'onoma de Puebla, Apartado Postal J-48, 72570 Puebla, M\'exico and}
\affiliation{Dual CP  Institute of High Energy Physics, Puebla,  M\'exico }
\affiliation{Facultad de Ciencias, Universidad de Colima, C.P. 28045, Colima, M\'exico}

\author{C. G. Honorato}
\email[]{carlosg.honorato@correo.buap.mx}
\affiliation{Facultad de Ciencias de la Electr\'onica, Benem\'erita Universidad Aut\'onoma de Puebla, Apartado Postal J-48, 72570 Puebla, M\'exico}

\author{S.~Moretti}
\email[]{stefano.moretti@cern.ch}
\affiliation{School of Physics and Astronomy, University of Southampton, Southampton, SO17 1BJ, United Kingdom}
\affiliation{Department of Physics and Astronomy, Uppsala University, Box 516, SE-751 20 Uppsala, Sweden}

\author{T.~Shindou}
\email[]{tetsuo.shindo@icu.ac.jp}
\affiliation{Division of Liberal-Arts, Kogakuin University,  2665-1 Nakano-machi, Hachioji, Tokyo, 192-0015, Japan}
\affiliation{Department of Natural Science, International Christian University, Osawa 3-10-2, Mitaka, Tokyo, 181-8585, Japan}

\date{\today}

\begin{abstract}
A $ Z_3 $ symmetric 3-Higgs Doublet Model (3HDM) with two inert doublets and one active doublet (which plays the role of the Higgs doublet),  the so-called I(2+1)HDM, is studied. We discuss the behaviour of this 3HDM realisation when one allows for a $ Z_3 $ soft-breaking term.  
In this setup, 
the lightest $Z_3$ charged neutral scalar  
can be a Dark Matter (DM) candidate.
If the breaking scale is small enough, this model provides a long-lived neutral state with the opposite CP parity to the DM state.
This long-lived particle could then be another effective DM candidate when its lifetime is comparable to the age of the universe. In this case, we have studied the properties of the ensuing relic density in the presence of the most recent limits from direct searches for DM. Conversely, the case in which the long-lived particle actually decays into the DM particle and Standard Model particles in a detector at a collider experiment is very attractive from the viewpoint of phenomenology.
Under such conditions, we have studied signatures involving missing transverse energy and multiple leptons (and jets) at the International Linear Collider (ILC) in the presence of displaced vertices.

\end{abstract}

\maketitle

\clearpage
\section{\label{introduction} Introduction}

A Higgs boson was discovered at the Large Hadron Collider (LHC) in July 2012 \cite{ATLAS:2012yve, CMS:2012qbp} 
and it has been shown that its nature is consistent with that of the Standard Model (SM), 
which contains only one SU(2) doublet field.
However, no compelling principle has ever been put forward that constrains the Higgs sector responsible for Electro-Weak Symmetry Breaking (EWSB) and mass generation to be one of the SM. 
In particular, there is no reason forbidding the introduction of new fields into the (pseudo)scalar sector of the underlying theory. 

There are many possibilities to extend the Higgs sector of the SM. As Nature seems to prefer doublet (pseudo)scalar fields, one could well restrict oneself to extensions of the SM that
only include such representations. Indeed, 
 the ensuing $N$-Higgs Doublet Model (NHDM) is a simple and attractive example, 
where the Higgs sector contains $N$ such fields.
The extension with doublet (pseudo)scalar fields also keeps $\rho=1$ at the tree level.
However, NHDMs generally lead to dangerous Flavour Changing Neutral Currents (FCNCs).
In order to suppress these, a discrete symmetry is often utilised. 
For example, in 2HDM, which is well studied in the literature, a softly broken $Z_2$ symmetry is usually introduced.
Under this $Z_2$ symmetry, one doublet is odd, and the other doublet is even.
Depending on the $Z_2$ parity assignment for the SM fermions, 2HDMs are then classified into four types \cite{Barger:1989fj,Grossman:1994jb,Aoki:2009ha}.

There are reasons to consider this kind of SM extension, as there are some problems that the SM cannot explain, calling for new physics. For example, within the SM, there is no viable candidate for Dark Matter (DM), there is no successful mechanism for baryogenesis, there is no dynamics that explains the smallness of the neutrino masses and so on.
One should then expect that Beyond the SM (BSM) scenarios can solve these problems and specifying the Higgs sector provides a vital clue to explore such new physics. NHDMs can, in particular, afford one with viable DM candidates.

For example, the so-called inert 2HDM, also called the Inert Doublet Model (IDM), provides one such DM candidate. 
In this model, even parity is assigned to the SM fermions, there is no softly broken term acting on the $Z_2$ symmetry in the Higgs potential and only the $Z_2$ even doublet gets a Vacuum Expectation Value (VEV) \cite{Deshpande:1977rw}.
Since the $Z_2$ is kept unbroken, the lightest $Z_2$ odd scalar is, therefore, stable and can be a DM if the particle is neutral. 

The inert sector of an NHDM with more than two doublets also provides DM candidates.
In the case of a 3-Higgs Doublet Model (3HDM), there are many possibilities to impose a discrete symmetry and trigger its breaking patterns. 
DM properties in an unbroken $Z_3$ symmetry model leading to two inert doublets are discussed in Ref.~\cite{Aranda:2019vda}.
If the CP symmetry is unbroken in the inert sector,  the lightest $Z_3$ charged particles in the CP-odd and CP-even sectors are individually stable. 
Thus, two DM candidates with opposite CP charges are provided.  This realisation is called a ``Hermaphrodite DM" scenario.

Although the ``Hermaphrodite DM'' scenario is 
attractive as a non-trivial multi-component DM scenario, 
it is difficult to prove it by 
collider experiments.
Because the two DM candidates are completely degenerate there.
Thus, we consider yet another DM scenario in the 3HDM inspired by  ``Hermaphrodite DM'', where 
a soft breaking term of the $Z_3$ symmetry is 
introduced.
In this scenario, there is no degeneracy between 
CP-even and CP-odd neutral scalars any more 
and only one scalar state is stable in the 
early universe,
while the $Z$ boson coupling with the DM candidate can be significantly suppressed 
in the same way as in the ``Hermaphrodite DM'' case, which may save the scenario from severe constraints stemming from Direct Detection (DD) experiments. Furthermore, 
owing to  the ensuing mass splitting between CP-even and CP-odd scalars, 
this scenario could also be explored by 
future collider experiments.

In this paper, we study phenomenological
probes when the model provides one DM candidate and one long-lived particle.
Since the two particles interact with the  $Z$, $W$ and Higgs bosons, we will prove that it should be possible to reveal both these two particles by isolating phenomenological properties for some processes leading to the same final state, proceeding through the two 
different states and pointing to their simultaneous presence. Furthermore, we show that when the mass difference is large enough for one of them to be unstable, an interesting phenomenology appears between the particles involved, where it is possible to obtain a spectacular signal of 6 leptons + missing transverse energy (or 4 leptons + 2 jets + missing transverse energy) by processes $e^+ e^- \to A_1 A_1 l^+ l^-$ followed of $A_1 \to DM + l^+ l^-$ (or $A_1 \to DM+ l^+ l^-$  and $A_1 \to DM + j j$) , in final states in a future ILC machine.

This paper is organised as follows.
Sec. II introduces the model with its like-Hermaphrodite DM scenario.
In Sec. III, we discuss the parameters used in our analysis. 
We show our numerical results in Sec. IV. 
Finally, we give our conclusions in Sec. V. 

\section{Model \label{model}}
\subsection{Lagrangian}
In this paper, 
we consider an extended Higgs sector with three Higgs doublets $\phi_i~(i=1,2,3)$.
We impose a $Z_3$ symmetry under which the three doublets transform as 
\begin{equation}
	\phi_1^{}\to \omega \phi_1^{}\;,\quad 
	\phi_2^{}\to \omega^2 \phi_2^{}\;, \quad 
	\phi_3^{}\to \phi_3^{}\;,
\end{equation}
with $\omega$ being a complex cubic root of unity, \textit{i.e.}, $\omega=e^{2\pi i/3}$.
The symmetric Higgs potential is given by 
\begin{equation}
	V=V_0+V_{Z_3}
\end{equation}
where $V_0$ is an invariant part under any phase rotation given by 
\begin{align}
	V_0 =& - \mu^2_{1} (\phi_1^\dagger \phi_1) -\mu^2_2 (\phi_2^\dagger \phi_2) - \mu^2_3(\phi_3^\dagger \phi_3) \\
	&+ \lambda_{11} (\phi_1^\dagger \phi_1)^2+ \lambda_{22} (\phi_2^\dagger \phi_2)^2  + \lambda_{33} (\phi_3^\dagger \phi_3)^2 \nonumber\\
	& + \lambda_{12}  (\phi_1^\dagger \phi_1)(\phi_2^\dagger \phi_2)
	+ \lambda_{23}  (\phi_2^\dagger \phi_2)(\phi_3^\dagger \phi_3) + \lambda_{31} (\phi_3^\dagger \phi_3)(\phi_1^\dagger \phi_1) \nonumber\\
	& + \lambda'_{12} (\phi_1^\dagger \phi_2)(\phi_2^\dagger \phi_1) 
	+ \lambda'_{23} (\phi_2^\dagger \phi_3)(\phi_3^\dagger \phi_2) + \lambda'_{31} (\phi_3^\dagger \phi_1)(\phi_1^\dagger \phi_3) \nonumber
	\end{align}
and $V_{Z_3}$ is a collection of extra terms ensuring the 
$Z_3$ symmetry given by 
\begin{equation}
	V_{Z_3} = \lambda_1(\phi_2^\dagger\phi_1)(\phi_3^\dagger\phi_1) + \lambda_2(\phi_1^\dagger\phi_2)(\phi_3^\dagger\phi_2) + \lambda_3(\phi_1^\dagger\phi_3)(\phi_2^\dagger\phi_3)  + \text{h.c.}
	\label{Z_3-3HDM}
\end{equation}

We adopt an ansatz that only $\phi_3$ has a VEV.
With this assumption, the EW symmetry is broken by $\langle \phi_3\rangle$ 
while the $Z_3$ symmetry is (initially) unbroken. 
A physical component in the $Z_3$ singlet field $\phi_3$ behaves like the SM Higgs boson, so we describe it as such using the label $h$.
Also, all SM particles have a $ Z_3 $ zero charge, so that only $ \phi_3 $ will couple to fermions. 
The Yukawa Lagrangian is given by
\begin{eqnarray}
	\mathcal{L}_{Y} &=& \Gamma^u_{mn} \bar{q}_{m,L} \tilde{\phi}_3 u_{n,R} + \Gamma^d_{mn} \bar{q}_{m,L} \phi_3 d_{n,R} \nonumber\\
	&& +  \Gamma^e_{mn} \bar{l}_{m,L} \phi_3 e_{n,R} + \Gamma^{\nu}_{mn} \bar{l}_{m,L} \tilde{\phi}_3 {\nu}_{n,R} + \text{h.c.}
\end{eqnarray}
Thanks to the $Z_3$ symmetry,  the lightest components of $\phi_1$ and $\phi_2$ can be stable and they both are DM candidates.
Given that $\phi_1$ and $\phi_2$ are inert, 
this model is termed I(2+1)HDM \cite{Ivanov:2012,Aranda:2019vda}.
Since the CP symmetry is also kept in the potential,  the combination of the CP and $Z_3$ symmetries  
predicts that these two DM candidates are such that one is CP-even and the other is CP-odd. However, evidently not being the real and imaginary part
of a complex field (as it will be clear below),  such a two-component DM is aptly named Hermaphrodite DM  \cite{Aranda:2019vda}.

As we will discuss later, the model with the softly broken term, only affecting the inert sector,   
\begin{equation}
	V_{\slashed{Z_3}}= -\mu_{12}^2(\phi_1^{\dagger}\phi_2)+\text{h.c.} 
\end{equation}
is the one interesting from the phenomenological point of view, though. 
 In fact, to realise proper EWSB, the parameter $ \mu_{12}^2 $ must be small, thus allowing for a $Z_3$ symmetry soft breaking. As a consequence, Although the stability of the DM candidates is affected, special cases can be found where two DM candidates are obtained. Indeed, if the aforementioned term is very small, we could approach the stability of both. One can show that a very long lifetime of the unstable particle is associated with the size of this parameter and its small size could make it a candidate for DM.

\subsection{Physical eigenstates}
The scalar potential acquires a minimum at the point
\begin{equation}
	\phi_1= 
	\left(\begin{array}{c}{\scriptstyle{H^{0+}_1}}\\ \frac{H^0_1+iA^0_1}{\sqrt{2}} \end{array}\right), \qquad \phi_2= 
	\left(\begin{array}{c}{\scriptstyle{H^{0+}_2}}\\ \frac{H^0_2+iA^0_2}{\sqrt{2}} \end{array}\right), \qquad \phi_3= 
	\left(\begin{array}{c}{\scriptstyle{H^{0+}_3}}\\ \frac{H_3^{0}+v+iA^0_3}{\sqrt{2}} \end{array}\right),
	\label{fields}
\end{equation}
where $v^2=\mu_{3}^{2}/\lambda_{33}$. 
Expanding the potential around the vacuum point we obtain the mass spectrum. In the model that allows a soft breaking of the $Z_3$ symmetry we have the following.
\begin{itemize}
	\item The neutral sector, CP-even scalars:
	\begin{eqnarray}
		&& \textbf{h} = H_3^{0} : \quad m^2_{h}= 2\mu_3^2 = 2 \lambda_{33} v^2.\\[1mm]
		&& \textbf{H}_1 = \cos\theta_h H^0_{1}+ \sin\theta_h H^0_{2}  
		\nonumber\\[1mm]
		&& \hspace{1cm}	
		m^2_{H_1}=  (-\mu^2_1 + \Lambda_{1})\cos^2\theta_h + (- \mu^2_2 + \Lambda_{2}) \sin^2\theta_h - (2\mu^2_{12} - \lambda_3 v^2) \sin\theta_h \cos\theta_h. 
		\nonumber\\[1mm]
		&& \textbf{H}_2 = -\sin\theta_h H^0_{1}+ \cos\theta_h H^0_{2} 
		\nonumber \nonumber\\
		&& \hspace{1cm}	
		m^2_{H_2}=  (-\mu^2_1 + \Lambda_{1})\sin^2\theta_h + (- \mu^2_2 + \Lambda_{2}) \cos^2\theta_h + (2\mu^2_{12} - \lambda_3 v^2) \sin\theta_h \cos\theta_h. 
		\nonumber
	\end{eqnarray}
	\item The neutral sector, CP-odd scalars:
	\begin{eqnarray}
		&& \textbf{A}_1 = \cos\theta_a A^0_{1}+ \sin\theta_a A^0_{2} \\[1mm]
		&& \hspace{1cm}	
		m^2_{A_1}= (-\mu^2_1 + \Lambda_{1})\cos^2\theta_a + (- \mu^2_2 + \Lambda_{2}) \sin^2\theta_a - (2\mu^2_{12} + \lambda_3 v^2) \sin\theta_a \cos\theta_a. 
		\nonumber\\[1mm]
		&& \textbf{A}_2 = -\sin\theta_a A^0_{1}+ \cos\theta_a A^0_{2}	\nonumber\\[1mm] && \hspace{1cm}		
		m^2_{A_2}= (-\mu^2_1 + \Lambda_{1})\sin^2\theta_a + (- \mu^2_2 + \Lambda_{2}) \cos^2\theta_a + (2\mu^2_{12} + \lambda_3 v^2) \sin\theta_a \cos\theta_a. \nonumber
	\end{eqnarray}
	\item The charged sector:
	\begin{eqnarray}
		&& \textbf{H}^\pm_1 = \cos\theta_c H^{0\pm}_{1} + \sin\theta_c H^{0\pm}_{2}\\[1mm]
		&& \hspace{1cm}	
		m^2_{H^\pm_1}= (-\mu^2_1 + \frac{1}{2}\lambda_{31}v^2)\cos^2\theta_c + (- \mu^2_2 + \frac{1}{2}\lambda_{23}v^2) \sin^2\theta_c - 2\mu^2_{12}\sin\theta_c \cos\theta_c. 
		\nonumber\\[1mm]
		&& \textbf{H}^\pm_2 = -\sin\theta_c H^{0\pm}_{1} + \cos\theta_c H^{0\pm}_{2}
		\nonumber\\[1mm]
		&& \hspace{1cm}	
		m^2_{H^\pm_2}= (-\mu^2_1 +\frac{1}{2}\lambda_{31}v^2)\sin^2\theta_c + (- \mu^2_2 + \frac{1}{2}\lambda_{23}v^2) \cos^2\theta_c + 2\mu^2_{12}\sin\theta_c \cos\theta_c. \nonumber
	\end{eqnarray}
\end{itemize}
Here, $\Lambda_1=(\lambda_{31}+\lambda^{\prime}_{31})v^2/2$ and $\Lambda_2=(\lambda_{23}+\lambda^{\prime}_{23})v^2/2$. The angles $\theta_h, \theta_a, \theta_c$ are the mixing angles for the scalar, pseudoscalar and charged mass-squared matrices, respectively, such that:
\begin{eqnarray}
	&& \tan 2\theta_c=\frac{4\mu_{12}^2}{2\mu_1^2-\lambda_{31}v^2-2\mu_2^2+ \lambda_{23}v^2}  
 =\varepsilon_c
 , \nonumber\\[1mm]
	&&\tan 2\theta_h=\frac{-\lambda_{3}v^2+2\mu_{12}^2}{\mu_1^2-\Lambda_1-\mu_2^2+ \Lambda_2} 
 , \nonumber\\[1mm]&& \tan 2\theta_a=\frac{\lambda_{3}v^2+2\mu_{12}^2}{\mu_1^2-\Lambda_1-\mu_2^2+ \Lambda_2}
 .
\end{eqnarray}
As we have mentioned above, 
in the case without the $Z_3$ breaking term, 
the CP symmetry makes $H_1$ and $A_1$ stable, so they can both be DM.
After EWSB, 
we have the vertex $ H_1A_1Z \propto \cos2\theta_h$, 
which leads to a too large cross-section for DM scattering
off nuclei and thus DD immediately 
excludes the scenario. 
In order to avoid this, 
the coupling constant of this vertex
should be significantly suppressed 
and this is realised by taking the condition of $\mu_1^2-\Lambda_1-\mu_2^2+ \Lambda_2=0$, so that 
$\theta_h=-\theta_a=\pi/4$ is satisfied.
As long as we keep this condition, 
the vertex $H_1A_1Z$ vanishes at tree-level, 
even when the $Z_3$ breaking term $\mu_{12}^2$ is taken into account.
However,  the vertex $H_1A_1Z$ can be induced at one-loop level, where the magnitude of this coupling is controlled by the soft-breaking term $\mu_{12}^2 $, hence, when the difference in mass between  $H_1$ and $A_1$ is small enough,  one could save this scenario from the strong DD limits. 
We show the analysis in the next section.

\section{Parameters for the analysis}
\subsection*{The input parameters}
As mentioned, in order to avoid the model being ruled out by DD bounds, we will consider the limit $ \theta_h = \pi / 4 $ where the masses squared can be written as follows: 
\begin{eqnarray}
	m_{H_1}^2&=&\frac{1}{2}(-\mu_1^2+\Lambda_1)+\frac{1}{2}(-\mu_2^2+\Lambda_2)-\frac{1}{2}(2\mu_{12}^2-\lambda_{3}v^2),\nonumber\\[1mm]
	m_{A_1}^2&=&m_{H_1}^2+2\mu_{12}^2,
 \label{m1}\\[1mm]
	m_{H_2}^2&=&\frac{1}{2}(-\mu_1^2+\Lambda_1)+\frac{1}{2}(-\mu_2^2+\Lambda_2)+\frac{1}{2}(2\mu_{12}^2-\lambda_{3}v^2),\nonumber\\[1mm]
	m_{A_2}^2&=&m_{H_2}^2-2\mu_{12}^2
 ,\label{m2}\\[1mm]
	m_{H_1^{\pm}}^2&=&-\mu_1^2+\frac{v^2}{2}\lambda_{31}-\frac{1}{2}\mu_{12}^2 \varepsilon_c,\nonumber\\[1mm]
	m_{H_2^{\pm}}^2&=&-\mu_2^2+\frac{v^2}{2}\lambda_{23}+\frac{1}{2}\mu_{12}^2 \varepsilon_c. \label{m3}
\end{eqnarray}
The input parameters $\lambda_{23}$, $\lambda_{13}$, $\lambda_{23}^{\prime}$, $\lambda_{31}^{\prime}$,  $\mu_1^2$ and $\mu_2^2$ can be rewritten in terms of the physical observables $m_{H_1}$, $m_{H_2}$, $m_{H_1}^{\pm}$, $m_{H_2}^{\pm}$, $\Lambda_{1}$ and  $\Lambda_{2}$. Introducing
\begin{eqnarray}
	&&\Delta_h=m_{A_1}-m_{H_1}\\ &&\Delta_c=m_{H_1^{\pm}}-m_{H_1},\\
	&&\delta_c=m_{H_2^{\pm}}-m_{H_1^\pm},\\
	&&\Delta_n=m_{A_2}-m_{H_1},  \label{Dn} \\ &&\Delta_n'=m_{H_2}-m_{A_1}, \\
	&&m_{H_2}^2= m_{A_2}^2+2\mu_{12}^2,
	\label{Delm}
\end{eqnarray}
\begin{eqnarray}
	g_1&=&\frac{g_{hH_1H_1}}{v}=\frac{1}{2}(\lambda_{23}+2\lambda_{3}+\lambda_{31}+\lambda_{23}^{\prime}+\lambda_{31}^{\prime}),\nonumber\\ g_2&=&\frac{g_{hH_1H_2}}{v}=\frac{1}{2}(\lambda_{23}-\lambda_{31}+\lambda_{23}^{\prime}-\lambda_{31}^{\prime}),
	\label{g12}
\end{eqnarray}
 where $g_{hH_1H_1}$ and $g_{hH_1H_2}$ are the coefficients of the vertices $hH_1H_1$ and $hH_1H_2$, respectively,
the Lagrangian parameters in terms of the observables reduce to:
\begin{eqnarray}
	\lambda_{23}&=&\frac{1}{v^2}\left((g_1+g_2)v^2-2m_{H_1}^{2}+2m^2_{H_2^{\pm}}\right) -\frac{2\mu_{12}^{2}}{v^2},\label{par1}\\[1mm]
	\lambda_{31}&=&\frac{1}{v^2}\left((g_1-g_2)v^2-2m_{H_1}^{2}+2m^2_{H_1^{\pm}}\right)-\frac{2\mu_{12}^{2}}{v^2},\\[1mm]
	\lambda_{23}^{\prime}&=&\frac{1}{v^2}\left(m_{H_1}^2+m_{H_2}^2-2m_{H_2^{\pm}}^{2}\right),\\[1mm]
	\lambda_{31}^{ \prime}&=& \frac{1}{v^2}\left(m_{H_1}^2+m_{H_2}^2-2m_{H_1^{\pm}}^2\right),\\[1mm]
	\lambda_3&=&\frac{1}{v^2}\left(m_{H_1}^{2}-m_{H_2}^2+2\mu_{12}^2 \right),\\[1mm]
	\mu_1^2&=&\frac{1}{2}\left((g_1 -g_2) v^2-2m_{H_1}^2\right)+\mu_{12}^{2},\\[1mm]
	\mu_2^2&=&\frac{1}{2}\left((g_1 +g_2) v^2-2m_{H_1}^2\right)+\mu_{12}^{2}.
\end{eqnarray}

\subsection*{Constraints on the model parameters}\label{constraints}
All  Benchmark Points (BPs) considered in this study  agree with the latest theoretical and experimental constraints that are applicable to the model, which are described in detail in \cite{Keus:2014isa,Aranda:2019vda}. For convenience, we recap these here.
As $\phi_3$ is identified with the SM Higgs doublet, $\mu_3$ and $ \lambda_{33}$ are Higgs field parameters and can be written in terms of the mass of the Higgs boson. We use the value $m_h=125$~GeV for the latter, so that 
\begin{equation} 
	m^2_h = 2\mu^2_3 = 2\lambda_{33} v^2.
\end{equation}
In agreement with perturbativity bounds and unitary conditions, we take the absolute values $|\lambda_i |\leq 3\pi$.
For the potential to be bounded from below, the following conditions are required 
\begin{eqnarray}
	&& \bullet ~ \lambda_{11}, \,\lambda_{22}, \,\lambda_{33} \geq 0, \\[1mm]
	&& \bullet ~  \lambda_{12} + \lambda'_{12}  + \sqrt{\lambda_{11}\lambda_{22}} \geq 0, \nonumber\\[1mm]
	&& \bullet ~ \lambda_{23} + \lambda'_{23} + \sqrt{\lambda_{22}\lambda_{33}} \geq 0,\nonumber\\[1mm]
	&& \bullet ~ \lambda_{31} + \lambda'_{31} + \sqrt{\lambda_{33}\lambda_{11}} \geq 0,\nonumber\\[1mm]
	&& \bullet ~ 
	\sqrt{\lambda_{11}\lambda_{22}\lambda_{33}} + (\lambda_{12} + \lambda'_{12}) \sqrt{\lambda_{33}} 
	+ (\lambda_{31} + \lambda'_{31}) \sqrt{\lambda_{22}} 	+ (\lambda_{23} + \lambda'_{23}) \sqrt{\lambda_{11}} \nonumber\\[1mm]
	&&
	+\sqrt{2 (\lambda_{12} + \lambda'_{12}  + \sqrt{\lambda_{11}\lambda_{22}})(\lambda_{23} + \lambda'_{23} + \sqrt{\lambda_{22}\lambda_{33}})(\lambda_{31} + \lambda'_{31} + \sqrt{\lambda_{33}\lambda_{11}})} \geq 0.
	\nonumber
\end{eqnarray}
For the $V_{Z_3}$ term not to dominate the behaviour of $V$, we also require the parameters of the $V_{Z_3}$ part to be smaller than the parameters of the $V_0$ part:
\begin{equation}
	|\lambda_1|, |\lambda_2|, |\lambda_3| < |\lambda_{ii}|, |\lambda_{ij}|, |\lambda'_{ij}| , \quad i\neq j = 1,2,3.
\end{equation}
Finally, as intimated, in $V_{\cancel{Z_3}} $, the parameter $\mu_{12}^2$ must be small. 

In our numerical studies, we have taken into account the following limits.
 In agreement with measurements done at LEP 	\cite{Cao:2007rm,Lundstrom:2008ai},  the limit on invisible decays of $ Z $ and $ W ^ \pm $ gauge bosons imply: 

\begin{equation}
m_{H_i^\pm} + m_{H_i,A_i} > m_{W^\pm} , \quad  m_{H_i} + m_{A_i} > m_Z, \quad 2m_{H_i^\pm} > m_Z . \end{equation}
Also,  the lower limit for the mass of the charged (pseudo) scalars is $m_{H^\pm_i} > 70-90 ~\mbox{GeV}.$
Furthermore, searches for charginos and neutralinos at LEP have been translated into limits on region of masses in the I(1+1)HDM \cite{Lundstrom:2008ai}, simultaneously requiring ($i=1,2$)
\begin{equation}
m_{H_i} \leq 80 ~ \mbox{GeV},  \quad m_{A_i} \leq 100 ~\mbox{GeV}\quad \mbox{and} \quad m_{A_i} - m_{H_i}  \geq  8~\mbox{GeV},  \nonumber
\end{equation}
otherwise a visible di-lepton or di-jet signal could appear.

The decay width of the SM-like Higgs boson into a pair of the inert scalars with $m_{S_i} < m_h/2$ is given by
\begin{equation}
	\Gamma (h\to S_iS_j) = \frac{g^2_{hS_iS_j} v^2}{32\pi m_h^3}
	\biggl[ \biggl(m_h^2-(m_{S_i}+m_{S_j})^2 \biggr)
	\biggl(m_h^2-(m_{S_i}-m_{S_j})^2 \biggr)
	\biggr]^{1/2},
	\label{Eq:Gamma_inv}
\end{equation}

where $S_i,S_j = H_1,A_1$, the coefficient $g_{hS_iS_j}\, v$ corresponds to the $hS_iS_j$ term in the Lagrangian and $m_{S_i}(m_{S_j})$ is the mass of the corresponding neutral inert particle $S_i (S_j)$. From the ATLAS experiment, it is possible to estimate the limit the SM-like Higgs boson invisible Branching Ratios (BR) as $\text{BR}(h\to \text{invisibles}) < 0.08-0.15.$ \cite{Aaboud:2019rtt} . Therefore, we have strong constraints on the {Higgs-DM} coupling. For our scenarios this BR is: 
\begin{equation}
	\text{BR}(h\to \text{invisibles}) = \frac{\Sigma_{i}\Gamma(h\to S_iS_i)}{\Gamma^{\text{SM}}_h + \Sigma_{i}\Gamma(h\to S_iS_i)},
	\label{Eq:BRinv}
\end{equation}
where $S_i =H_1, \, A_1$.  Due to the constraints coming from $h\to\gamma\gamma$,  the parameters and the inert masses in our model  are in agreement with experiments as the one presented in \cite{Cordero-Cid:2018man}.
Regarding constraints coming from $h\to\gamma\gamma$, the inert charged masses and parameters in our analysis fall within the acceptable ranges obtained in Ref.~\cite{Cordero-Cid:2018man} where a combined ATLAS and CMS Run 1 limit was used for the SM-like Higgs signal strengths.

\subsection*{Constraints  from experimental data on DM}

The relic density constraint, according to the last measurements from the Planck experiment \cite{plank2020}, is:  
\begin{equation*}
\Omega_{\rm DM}h^2 = 0.120 \pm 0.001.
\end{equation*}
The DD results on DM measured in XENON1T \cite{Aprile:2018dbl} and LUX-ZEPLIN 2023 \cite{LZ:2022lsv} are used. The Indirect Detection (ID) results of FermiLAT \cite {Karwin:2016tsw} are also adopted (these strongly restrict the annihilation of DM in the final states $ b \bar b $ and $ \tau^+ \tau^- $).   However, the presence of the vertex $Z H_1 A_1$ modifies the previous analysis, where the hermaphrodite model is presented with $Z_3$ exact symmetry \cite{Aranda:2019vda}. Thus, firstly we should analyse the aforementioned coupling and its consequences on the particle $A_1$, although it could be unstable, it could have a long lifetime close to the stage of the early universe.

\subsection*{The vertex $Z H_1 A_1$ inducing the decay  $\Gamma(A_1\to H_1 f \bar f)$ at one-loop level}

\begin{figure}[!t]
\begin{center}
\includegraphics[scale=.4]{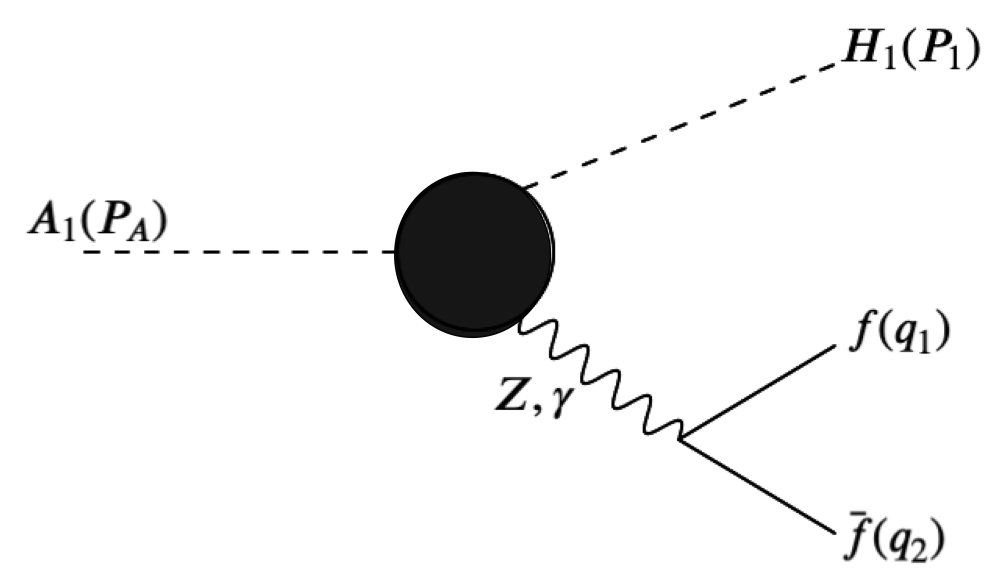} 
\caption{General structure for reducible diagram, the momentum configurations are $P_A=P_1+q_1+q_2$.}
\label{reducible}
\end{center}
\end{figure}

As we mentioned,   the vertex $ H_1A_1Z $ must not lead to a too large cross-section for DM scattering off nuclei, so we should understand where this coupling is permitted by DD. Thus, we proceed to analyse it. In our model,  despite the vertex $ ZH_1A_1$ being absent at tree-level, it can be induced at one loop-level, through the decay $~A_1\to ~H_1 f \bar{f} $ being kinematically permitted when $m_{A_1} - m_{H_1} > 2 m_{f}$.   The width $\Gamma(A_1\to H_1 f \bar f)$ receives contributions from irreducible and reducible diagrams. The irreducible diagrams consist of diagrams with $W^\pm$ bosons, charged (pseudo)scalars and fermions into the loop. Furthermore, the reducible contribution is introduced via the vertex  $A_1 H_1 Z^*$, which involves also neutral (pseudo)scalar states in the loops. 
Firstly, we analyse the reducible diagrams (see Figure \ref{reducible}). This set of diagrams has the following contribution:
\begin{equation}
  A_1(P_A)\to H_1(P_1) Z^*(P_A-P_1) \to H_1(P_1) f(q_1)\bar f(q_2).
\end{equation}  
In this way, the amplitude arising from the reducible diagrams can be written as follows:
\begin{equation}
{\cal {M}}_R=\bar v(q_1)\gamma^\nu(g_Z^V-g_Z^A \gamma^5)u(q_2) 
\left[\frac{g_{\mu\nu}-\frac{(q_1+q_2)_\mu (q_1+q_2)_\nu}{m_Z^2}}{(q_1+q_2)^2-m_Z^2}\right]
g_{A_1H_1Z^*},
\end{equation}
where $\gamma^\nu(g_Z^V-g_Z^A\gamma^5)$ represents the general vertex $Zf\bar f$. The propagator for the $Z$ boson is introduced in unitary gauge, and $g_{A_1H_1Z}$ is the coupling function at one loop level, given by:
\begin{equation}
g_{A_1H_1Z}=i\mu_{12}^2[k(P_A+P_1)^\mu+k^\prime (P_A-P_1)^\mu],
\end{equation}
    where $\mu_{12}^2$ is the $Z_3$ soft-breaking symmetry parameter. For the case with $Z_3$ exact symmetry, the vertex $g_{A_1H_1Z}$ is identically zero at any order of radiative corrections. Furthermore, the momentum configuration in accordance with Figure \ref{reducible} is: 
\begin{equation}
P_A^\mu-P_1^\mu=q_1^\mu+q_2^\mu.
\end{equation}
Using the Dirac equation for light fermions, one gets: 
\begin{equation}
\bar v(q_1)(\slash \hspace{-1.6 mm} q_1+\slash \hspace{-1.6 mm} q_2)u(q_2)\approx 0.
\end{equation}
The effective vertex can be written as:
\begin{equation}\label{effective}
g_{A_1H_1Z^*}=i\mu_{12}^2k(P_A^\mu+P_1^\mu).
\end{equation}
For the vertex $A_1H_1 Z^*$ at one loop level, in the end, we have found that the main contributions are given by neutral scalar (vector) bosons, which in the following sections are discussed.  

\subsection{Charged bosons contributions}

For the case where we consider the sector of charged scalar and vector bosons, the couplings involved in the loop calculation have the same structure as the vertex ($A^\mu H_1 A_1$), which is absent due to CP conservation, since the couplings are of the Electro-Magnetic (EM) type, expressed by the following interactions:
\begin{eqnarray}
g_{Z H_i^+ H_i^-}&=&\frac{1-2s_W^2}{2c_W s_W} g_{\gamma H_i^+H_i^-}, \\
g_{Z W^+W^-}&=&\frac{c_W}{s_W}g_{\gamma W^+W^-},\\
g_{ZW^+H_i^- A_1}&=& \frac{-s_W}{c_W}g_{\gamma W^+H_i^- ~A_1},\\
g_{ZW^+H_i^- H_1}&=& \frac{-s_W}{c_W}g_{\gamma W^+H_i^- H_1}.
\end{eqnarray}
From these relations, one can see that this loop contribution is absent as well due to CP conservation (see \cite{Cordero:2017owj}). 

\subsection{Neutral boson contributions}

\begin{figure}
\centering 
\includegraphics[scale=.04]{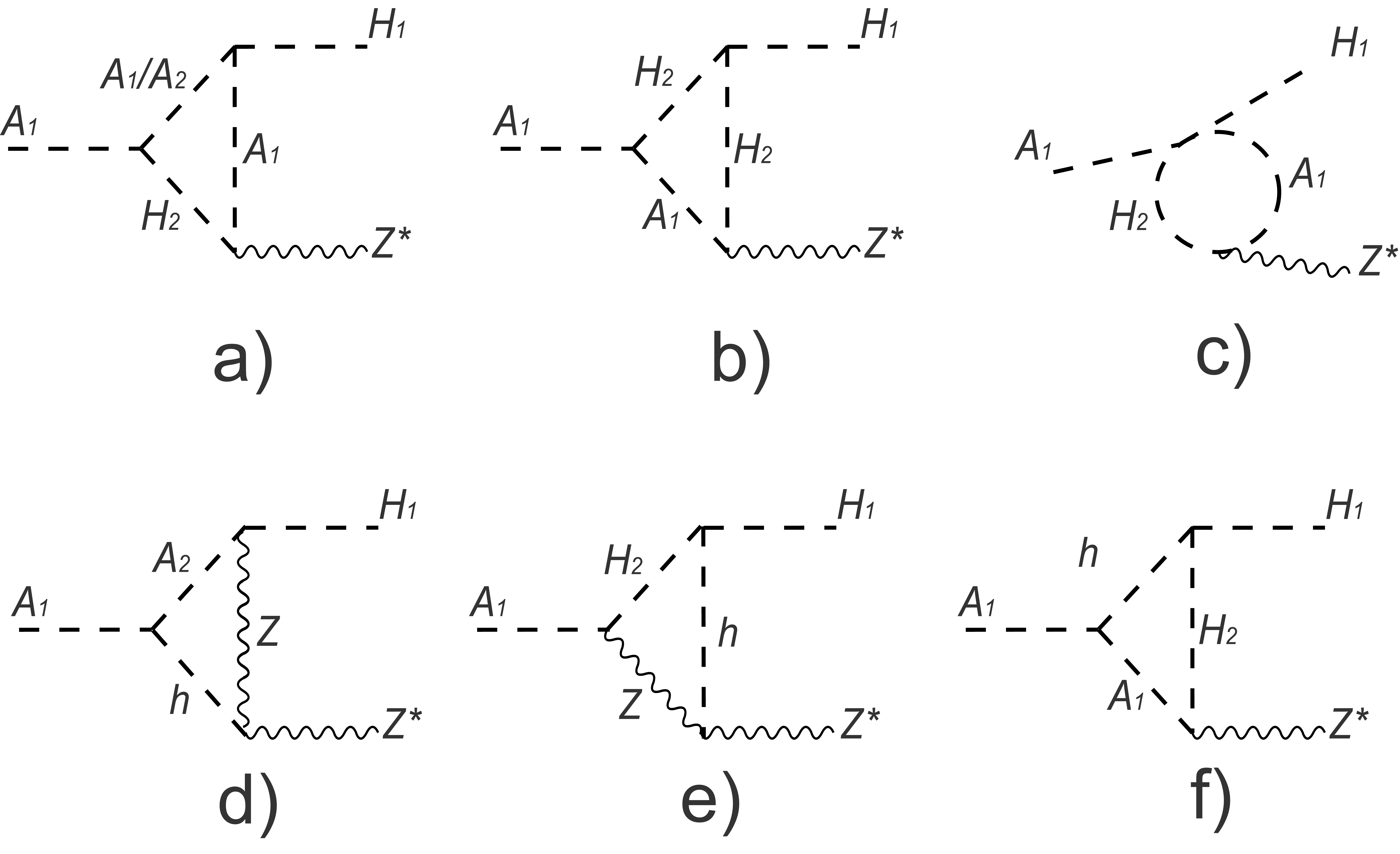} 
\caption{Set of diagrams with neutral contributions. The diagrams in the top panel represent a subset whose contribution is negligible. In contrast, the bottom subset has relevant contributions. } 
\label{sneutro}
\end{figure}
Now, we introduce the contributions of the neutral scalars and the $Z$ boson into the loops. These are given by the diagrams in Figure \ref{sneutro}. For the loop configurations shown in the top panel of Figure \ref{sneutro}, the amplitudes from diagrams a),  b) and c) are given by: 
\begin{eqnarray}
{\cal M}_{a}^\mu(m_{A_i}) &=&\frac{-\lambda_{A_i} i e (\lambda_1^2-\lambda_2^2) v^2 }{8s_{2W}}\nonumber\\
&&\times \int\frac{d^nq}{(2\pi)^n}\frac{(P_1^\mu+P_A^\mu+2q^\mu)}{(q^2-m_{A_i}^2)[(P_1+q)^2-m_{A_1}^2][(P_A+q)^2-m_{H_2}^2]},
\end{eqnarray}
\begin{eqnarray}
{\cal M}_{b}^\mu &=&\frac{i e (\lambda_1^2-\lambda_2^2) v^2 }{8s_{2W}}\nonumber\\
&&\times \int\frac{d^nq}{(2\pi)^n}\frac{(P_1^\mu+P_A^\mu+2q^\mu)}{(q^2-m_{H_2}^2)[(P_1+q)^2-m_{H_2}^2][(P_A+q)^2-m_{A_1}^2]},
\end{eqnarray}
\begin{equation}
{\cal M}_{c}^{\mu}=\frac{-i e(\lambda_{11}-\lambda_{22})}{4s_Wc_W} \int\frac{d^nq}{(2\pi)^n}\frac{(P_1^\mu-P_A^\mu+2q^\mu)}{(q^2-m_{A_1}^2)[(P_1-P_A+q)^2-m_{H_2}^2]},
\end{equation}
where $\lambda_{A_1}=3$ and $\lambda_{A_2}=1$ for each diagram from a).  Nevertheless, notice that the contributions given by {a)} and {b)} in Figure \ref{sneutro} are negligible because they are proportional to the difference $\lambda_1^2-\lambda_2^2$ (in our numerical analysis, these two parameters are equal to each other.). On the other hand,  the contribution of diagram {c)} in Figure \ref{sneutro}  is proportional to $(P_A^\mu-P_1^\mu)=(q_1^\mu+q_2^\mu)$, which is also negligible according to the Dirac equation for light fermions. Therefore, the contributions from the three diagrams in the top panel of Figure \ref{sneutro} can be neglected.
Thus, the vertex $g_{A_1H_1Z}$ is given by the contributions from the diagrams in the bottom panel in Figure \ref{sneutro}. The contributions {d)} and {e)} are proportional to $g_2$, which appears in the couplings for $hA_1A_2$ and $hH_1H_2$. In order to cancel divergences, both diagrams are computed together (see Eq. \ref{g2}). In contrast, the contribution of the diagram {f)} is proportional to $g_2 g_1$ (due to the product of couplings $hH_1H_2$ and $A_1A_1h$). 
Besides, this last diagram is divergence-free by itself (see Eq. \ref{g1g2}). 
Now, we present the explicit contribution of these diagrams, 
\begin{eqnarray}
{\cal M}_{d}^\mu &=&\frac{i e^3g_2v^2}{4s_W^3c_W^3}\int\frac{d^n q}{(2\pi)^n}\frac{(q^\alpha-P_1^\alpha)\left(g^{\alpha\mu}-\frac{(P_1+q)^\alpha(P_1+q)^\mu}{m_Z^2}\right)}{(q^2-m_{A_2}^2)[(P_1+q)^2-m_Z^2][(P_A+q)^2-m_h^2]},\\
{\cal M}_{e}^\mu &=&\frac{i e^3g_2v^2}{4s_W^3c_W^3}\int\frac{d^n q}{(2\pi)^n}\frac{(q^\alpha-P_A^\alpha)\left(g^{\alpha\mu}-\frac{(P_A+q)^\alpha(P_A+q)^\mu}{m_Z^2}\right)}{(q^2-m_{H_2}^2)[(P_1+q)^2-m_h^2][(P_A+q)^2-m_Z^2]},\\
{\cal M}_{f}^\mu &=&\frac{i e g_1 g_2 v^2 }{2c_W s_W}\int\frac{d^nq}{(2\pi)^n}\frac{(P_1^\mu+P_A^\mu+2q^\mu)}{(q^2-m_h^2)[(P_1+q)^2-m_{H_2}^2][(P_A+q)^2-m_{A_1}^2]}. 
\end{eqnarray}
Using FeynCalc and LoopTools for analytical and numerical evaluation \cite{Mertig:1990an,Hahn:1998yk}, respectively, we obtain the effective coupling of the vertex $A_1H_1Z$  \ref{effective}
\begin{equation}
g_{A_1H_1Z^*}=i\mu_{12}^2k(P_A^\mu+P_1^\mu)=i\mu_{12}^2(k_{d+e}+k_f)(P_A^\mu+P_1^\mu)=ig_3(P_A^\mu+P_1^\mu)\\
\end{equation}
being 
\begin{eqnarray}
{\cal{M}}_{d}^\mu +{ \cal{M}}_{e}^\mu&=&i\mu_{12}^2 k_{d+e}(P_A^\mu+P_1^\mu), \\ 
{\cal{M}}_{f}^\mu &=&i\mu_{12}^2 k_{f}(P_A^\mu+P_1^\mu),
\end{eqnarray}
where $k_{d+e}$ and $k_f$ are given by  Passarino-Veltman scalar functions 
through the following expressions \cite{Passarino:1978jh}: 
\begin{eqnarray}
k_{d+e}&=&\frac{-i e^3v^2g_2}{128 \pi^2 s_W^3c_W^3 m_Z^2[(m_{12}^2-2\mu_{12}^2)^2-4m_{H_1}^2m_{12}^2]}\nonumber\\
&&\times\Big\{\Delta_Z^{+-}\Big[R_{12}(\Delta B^{A_1,A_2,h}_{m_{12},h,Z}+\Delta B^{A_1,H_2,Z}_{m_{12},h,Z}+\Delta B^{H_1,A_2,Z}_{m_{12},h,Z}+\Delta B^{H_1,h,H_2}_{m_{12},h,Z})\nonumber\\
&&+2(\Delta B^{A_1,A_2,h}_{H_{1},h,H_2}+\Delta B^{A_1,H_2,Z}_{H_1,A_2,Z})\Big]+\Big[m_{12}^2R_{12}\Delta_Z^{--}+R_{12}[\Delta_{12}^-(\delta_{12}^-\nonumber\\
&&+2\Delta_{12}^-+\Delta_h^{+-})+m_Z^4+m_Z^2(\delta_h^{-+}-2\delta_{H_1}^{++})]+2\big(-\Delta_{12}^-m_h^2+m_Z^4\nonumber\\
&&+m_Z^2(2\delta_{12}^-+4\Delta_{12}^--\Delta_h^{+-})\big)\Big]C_{ZA_2h}+
\Big[m_{12}^2R_{12}\Delta_Z^{--}\nonumber\\
&&+R_{12}[\Delta_{12}^-(\Delta_h^{+-}-\delta_{12}^-)+m_Z^4-m_Z^2(2\delta_{H_1}^{++}-\delta_h^{-+}+4\mu_{12}^2)]\nonumber\\
&&+2\big(\Delta_{12}^-m_h^2-m_Z^4+m_Z^2(2\delta_{12}^-+\Delta_h^{+-})\big)\Big]C_{hH_2Z}\Big\}, \label{g2}
\end{eqnarray}
and
\begin{eqnarray}
k_f&=&\frac{iev^2 g_1 g_2}{128\pi^2s_Wc_W[(m_{12}^2-2\mu_{12}^2)^2-4m_{H_1}^2m_{12}^2] }\Big\{2\Delta B^{H_{1},h,H_2}_{A_1,A_1,h} \nonumber\\
&&+	R_{12}\Big(\Delta B^{m_{12},A_1,H_2}_{H_1,h,H_2}+\Delta B^{m_{12},A_1,H_2}_{A_1,A_1,h}\Big)
+\Big(m_{12}^2R_{12} \nonumber\\
&&-R_{12}(\delta_{12}^++4\mu_{12}^2-2m_h^2+2m_{H_1}^2)+2\Delta_{12}^-\Big)C_{H_2hA_1}, \label{g1g2}
\end{eqnarray}	
where
\begin{eqnarray}
\Delta B_{d,e,f}^{a,b,c}&=&B_0[m_a^2,m_b^2,m_c^2]-B_0[m_d^2,m_e^2,m_f^2],\\
C_{abc}&=&C_0[m_{H_1}^2,m_{A_1}^2,m_{12}^2,m_a^2,m_b^2,m_c^2],\\
R_{12}&=&m_{12}^2/\mu_{12}^2,\\
\delta_{12}^\pm &=& m_{H_2}^2\pm m_{H_1}^2,\\
\delta_a^{\pm\mp} &=& m_a^2\pm\delta_{12}^\mp,\\
\Delta_{12}^\pm&=&2\mu_{12}^2\pm \delta_{12}^-,\\
\Delta_a^{\pm\mp}&=&m_a^2\pm\Delta_{12}^{\mp}.
\end{eqnarray}
In Eqs. \ref{g2}--\ref{g1g2}, we also have used the following relations between the masses:
\begin{eqnarray}
&&m_{A_1}^2=m_{H_1}^2+2\mu_{12}^2,\\
&&m_{A_2}^2=m_{H_2}^2-2\mu_{12}^2.
\end{eqnarray}
Finally, the  one-loop contributions are given in terms of $Z_3$ soft-breaking term ($\mu_{12}^2$) and the kinematic variable $m_{12}^2=(P_A-P_1)^2$. 
 In short, the decay width $\Gamma(A_1\to ~H_1 f \bar{f})$ (computed at one-loop level)  could be associated to a very long lifetime (even around the universe age) when $\mu_{12}^2$ is small enough.

\subsection*{Parameter scan and benchmark points}
\begin{figure}
\centering
\includegraphics[width=8cm, height=6cm]{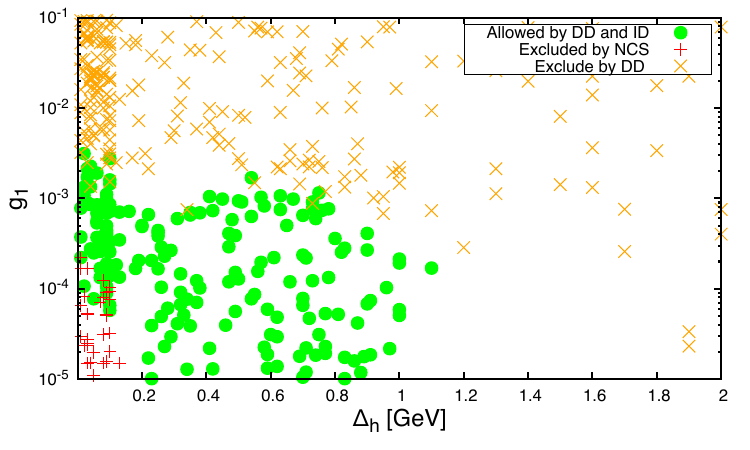} 
\includegraphics[width=8cm, height=6cm]{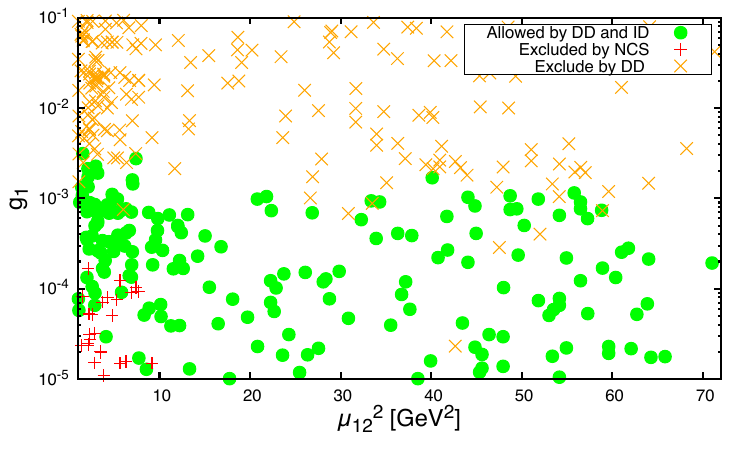} 
\caption{Set of values for $g_{1}$ against $\Delta_h$ (left) and $\mu_{12}^2$ (right).  Red dots are forbidden by Neutrino Coherent Scattering (NCS).  The orange dots are above the experimental limit imposed by LZ.  The green dots satisfy relic density, as well as DD and ID experimental data.}
\label{g1}
\end{figure}

\begin{figure}
\centering
 \includegraphics[width=8cm, height=6cm]{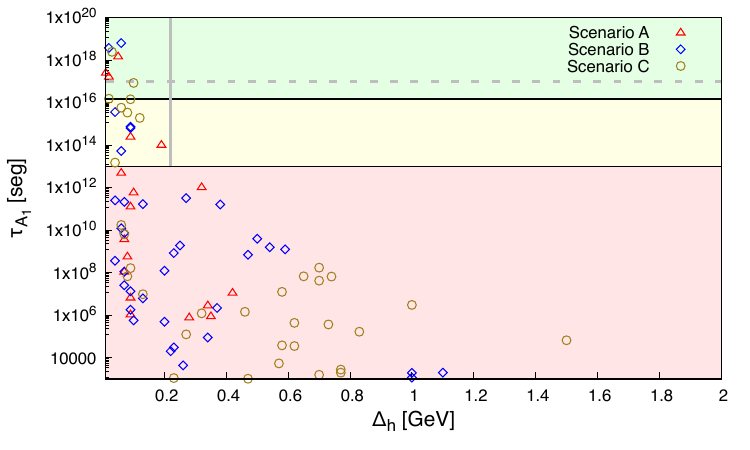}
 \includegraphics[width=8cm, height=6cm]{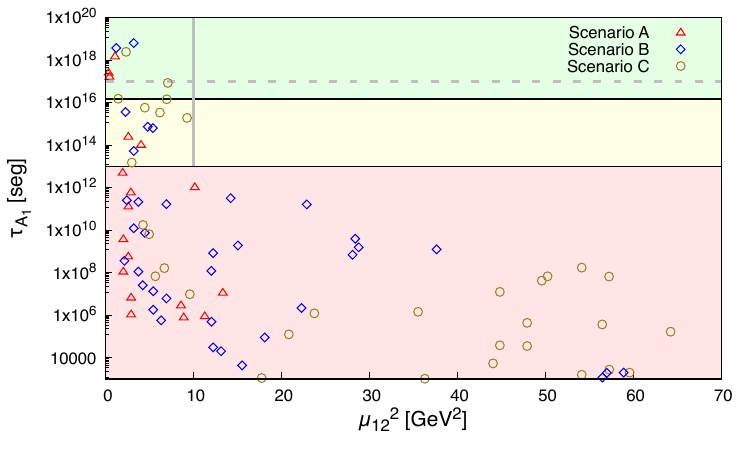}
\caption{Lifetime for DD accepted points versus $\Delta_h$ (left) and $\mu_{12}^2$ (right). The green region represents lifetimes around the universe age and time of formation of the youngest galaxies (the horizontal dashed line represents the universe age). The yellow zone depict a long lifetime particles which could contribute to DM. The red region contains BPs that could affect Big Bang Nucleosynthesis (BBN).}
\label{time1}
\end{figure}

As we discussed previously, for  the $H_1$ and $A_1$ particles to qualify as viable DM candidates, the  $ H_1A_1Z$ vertex 
must vanish at tree-level and only be induced at the loop level. Therefore, $\theta_h = \pi/4$ is the only acceptable value in the $0 \leq \theta_h < \pi$ range and $\mu_1^2-\Lambda_1-\mu_2^2+ \Lambda_2=0$ must be satisfied for the model to qualify as a viable DM framework.  Specifically, the constraints on the parameters,  according to the invisible Higgs decay rates, require that the coefficient of the $hH_1H_2$ vertex must be  $-0.029 \leq g_1 \leq 0.029$. Assuming the parameter relations given in  Eqs. (\ref{m1})--(\ref{m3}) and (\ref{g12})--(\ref{par1}),  we first consider DD and ID constraints on DM as well as theoretical ones. 
\begin{table}[h]
\centering
\begin{tabular}{|l|l|}
\hline 
Scanned parameters & Fixed parameters \\ 
\hline 
$10\ {\rm GeV} \leq M_{h_1}\leq 80 \ {\rm GeV}$ & $\lambda_1=\lambda_2=\lambda_{12}'=0.1$\\ 
$1\ {\rm GeV} \leq \Delta_n \leq 50\ {\rm GeV}$ & $\lambda_{11}=\lambda_{22}=\lambda_{12}=0.1$\\ 
$1\ {\rm GeV} \leq \Delta_c \leq 50\ {\rm GeV}$ & $\delta_c=10$ \\ 
$10^{-5}\leq g_{1},g_2\leq 10^{-1}$& $\Delta_h=\{0.01,5\}$ GeV\\
$10^{-9}\leq g_3 \leq 10^{-3}$ & \\
\hline 
\end{tabular} 
\caption{Parameter values for the scanning process in microMEGAs.} 
\label{parameters}
\end{table}
Figure \ref{g1}  shows a random scan of the allowed values for parameters space of the model given in Table \ref{parameters}, one can constraint even more the parameter $g_1$, the allowed values are $ 10^{-5} \leq g_1\leq 10^{-3}$ for 0.2 GeV $\leq \Delta_h \leq 2$ GeV and 10 GeV$^2$ $\leq \mu_{12}^2 \leq 70$ GeV$^2$. In Figure \ref{g1}, we show a set of points permitted over the parameter space of our model, given by the green symbols. 
\begin{figure}
\centering
\includegraphics[scale=1]{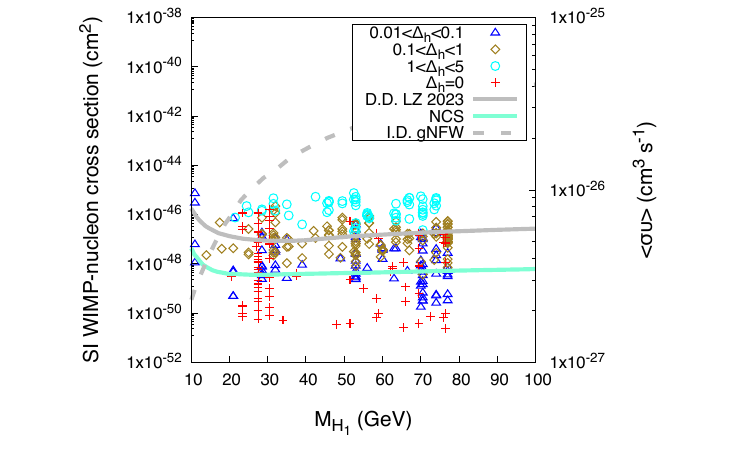} 
\caption{Spin Independent (SI) DM-nucleon cross-section as a function of the $H_1$ mass for different values of $\Delta_h$. The lines represent: (dashed) upper limit for ID, (continuous grey) upper limit for DD \cite{LZ:2022lsv} and (continuous blue) lower limit for Neutrino Coherent Scattering (NCS). In the right vertical axis, $<\sigma v>$ represents the thermally averaged cross-section.  }
\label{micro}
\end{figure}
Taking into account the aforementioned scanning, we get from micrOMEGAs the results of  Figure \ref{micro}, where relic density, DD and ID experimental constraints are considered. One can see that the points enclosed by solid lines, brown and blue, as well as the dashed grey line, are permitted. We show several allowed cases for  0.1 GeV $\leq \Delta_h \leq 1$ GeV (brown circle) and  0.01 GeV $\leq \Delta_h \leq 0.1$ GeV (blue triangle), as well as the exact $Z_3$ symmetry case $\Delta_h=0$ (red cross). We have considered the one-loop $A_1H_1Z$ vertex contribution, which is implemented  at microMEGAs 5.2.4 \cite{Belanger2015} as an effective vertex.   All set of points shown in Figure \ref{micro} satisfy the relic density,  but only a few points satisfy the DD constrains from LUX-ZEPLIN 2023 \cite{LZ:2022lsv}.  In fact,  for $\Delta_h=1$ GeV, there are no acceptable points.  However, for smaller values of $\Delta_h$, it is possible to define three relevant scenarios:
\begin{itemize}
    \item Scenario A: $20~{\rm GeV}< m_{H_1} <35 ~{\rm GeV}$
     \item Scenario B: $50~{\rm GeV}< m_{H_1} <65 ~{\rm GeV}$
      \item Scenario C: $70~{\rm GeV}< m_{H_1} <80 ~{\rm GeV}$
\end{itemize}
Besides, considering the lifetime of the particle $A_1$ in our model, we could have the following $\mu_{12}^2$ stages in the model: 
\begin{enumerate}
    \item
    One can get a feasible two-component DM scenario when $\mu_{12}^2$ is smaller than $\mathcal{O}(0.1)$~GeV$^2$, when one inert state is stable DM ($H_1$) and another inert particle ($A_1$) could be considered a second DM candidate due to its long lifetime.
    
    \item 
    We have only one-component DM, $H_1$, and  $A_1$ is unstable, but then it can decay inside the detector for 3 GeV $\lesssim \Delta_h \lesssim  6$ GeV.
\end{enumerate}
Considering DM constraints for stage 1, the prediction of the total relic density due to the presence of both $H_1$ and  $A_1$ is given by $\Omega_{\rm DM}h^2=\Omega_{H_1}h^2+\Omega_{A_1}h^2$. In stage 2, $\Omega_{\rm DM}h^2=\Omega_{H_1}h^2$.
Considering the aforementioned scenarios,  we have selected our BPs. 
In practice, we choose parameter space configurations where the lifetime of the particle
$A_1$ could be long enough that it can be a second particle candidate for DM. 
In Figure \ref{time1} the behaviour of lifetime versus $\Delta_h$ and $\mu_{12}^2$ is presented.  One can see that the vertical grey line encloses the points where both DM states are possible, taking into account both green and yellow regions.
Here, we present the analysis of a set of BPs, combining the results of Figures \ref{g1} and \ref{micro}.
We select BPs with lifetimes over $10^{14}$~sec, corresponding to the dark age of the universe: in this connection, the preferred values for $m_{H_1}$ are between $60-63\ {\rm GeV}$ (Scenario B) and $77-78\ {\rm GeV}$ (scenario C). 
In Scenario B, the optimal values for  $\Delta_n=m_{H_2}-m_{H_1}$ are between $10-30\ {\rm GeV}$. In contrast, for scenario C, the preferred range is $\Delta_n$ between $5-25 \ {\rm GeV}$. 
The coupling factor $g_1$ is 
between $0.0002$ and $0.0003$ for Scenario B  and between $0.0005$ and $0.002$ for scenario C. About the second coupling factor, $g_2$, the preferred range is $0.005$--$0.02$ for both scenarios B and C. 
We can see a saturation of the relic density in the mass ranges  $ 52 $ GeV $\leq m_{\rm H_1} \leq 65  $ GeV and $ 75 $ {\rm GeV}$\leq m_{\rm H_1} \leq 78$ GeV.
The values of the coupling factor at one loop, $g_3$, is within the range of $10^{-7}-10^{-6}$ for $\Delta_h<0.05\ {\rm GeV}$ in both scenarios. 
 \begin{figure}
\centering
  \includegraphics[width=8cm, height=6cm]{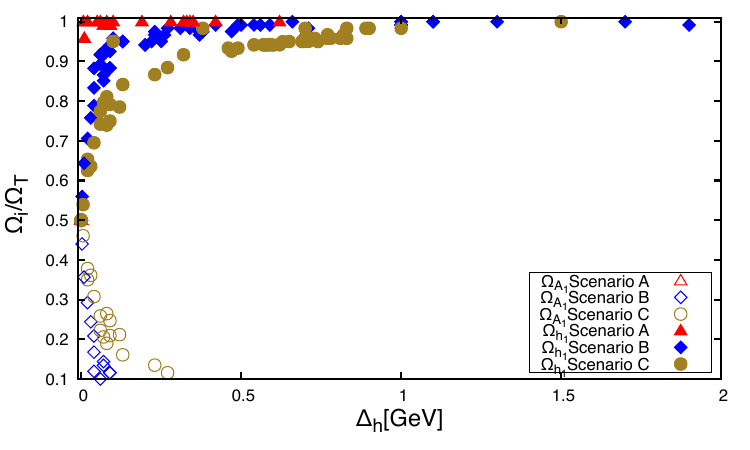}
  \includegraphics[width=8cm, height=6cm]{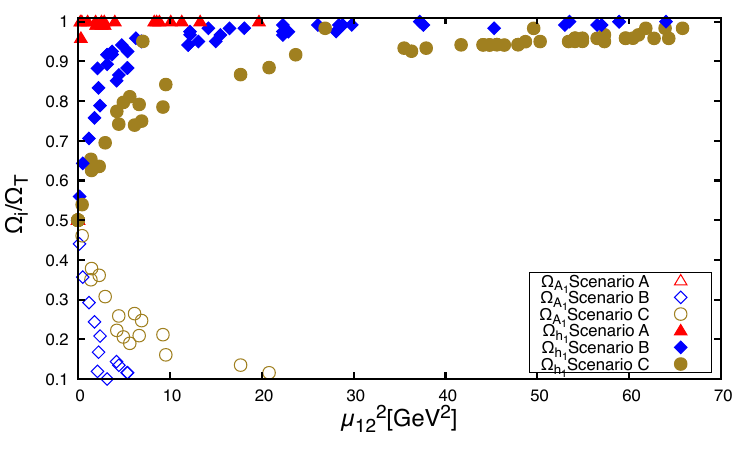}
\caption{Ratio  of the contribution from $H_1$ and $A_1$ to the total relic density. The hollow points represent the possibility of an $A_1$ DM-like candidate with long lifetime, while the filled points correspond to the case when the relic density is satisfied for $H_1$, via $\Omega_{H_1}^2$, which would be the sole DM candidate at any time. }
\label{time2}
\end{figure}

For the numerical evaluation of the DM abundance, we have used micrOMEGAs 5.2.4 \cite{Belanger2015}, which produced the results in
Figure \ref{time2}, showing a scan of the combined relic density of the two components of DM, $H_1$ and $A_1$, for the parameters shown in Figure \ref{micro}, considering the scenarios A, B and C. 
The filled points are associated with $\Omega_{H_1}/\Omega_{T}$ and the hollow points represents the ratio $\Omega_{A_1}/\Omega_{T}$, where $\Omega_{T}$ is the total density, $\Omega_{T}=\Omega_{H_1}+\Omega_{A_1}$. 
We can get the same relic density for both $A_1$ and $H_1$, when the $Z_3$ symmetry is exact ($\Delta_h = 0$ or $\mu_{12}^2=0$). 
When the parameters $\Delta_h $ and/or $\mu_{12}^2$ increase, the relic density decreases for $A_1$, leaving $H_1$ to saturate the entire relic density.
For $m_{H_1}\sim m_h/2$, the $s$-channel exchange of the SM-like Higgs boson is efficient in the annihilation of $H_1$ and $A_1$, whereas,   
for $m_{H_1}>75$ GeV, the process
$H_1H_1\to WW^*$ dominates the annihilation channels and is just as efficient. 
Therefore, in such regions, we can avoid 
overproduction of DM. 
In   Table \ref{results1}, we summarise our BPs, also considering the coupling strength $g_3$ of the coupling $A_1H_1Z^*$. The value of $g_3$ depends on the parameter $\mu_{12}^2$ or its associated parameter $\Delta_h$. For each value of $g_3$ and $\Delta_h$, we
 use CalcHEP \cite{Belyaev:2012qa} to calculate the total width and lifetime of $A_1$. 
\begin{table}[h] 
\centering
\begin{tabular}{|c|c|c|c|c|}
\hline 
$\tau c$ [m]&$\tau~[{\rm s}]$ &$g_3$& $\Delta_h~[{\rm GeV}]$ &  $\mu_{12}^2~[{\rm GeV}^2]$  \\ 
\hline
\hline
\multicolumn{5}{|c|}{Scenario A}\\
\hline
 $1.9\times 10^{-3}$&$6.5\times 10^{-12}$& $5.1\times 10^{-5}$ & 10
 & 370\\ \hline
 $4.05$&$1.3\times 10^{-9}$& $2.5\times 10^{-5}$ & 5 & 172\\ 
 \hline
 $2.8\times 10^3$&$9.4\times 10^{-5}$& $1.5\times 10^{-5}$ & 3 & 105\\ \hline
 $4.6\times 10^5$&$1.5\times 10^{-3}$& $1\times 10^{-5}$ & 2 & 66\\ 
 \hline
 $2.7\times 10^{8}$&$9.1\times 10^{-1}$& $5.1\times 10^{-6}$ & 1 & 32.5\\ \hline
 $5.2\times 10^{11}$&$1.7\times 10^3$& $2.5\times 10^{-6}$ & 0.5 & 16.1\\ \hline
 $8.9\times 10^{17}$&$2.9\times 10^{9}$& $1\times 10^{-6}$ & 0.2 & 6.4\\ \hline
 $4.6\times 10^{20}$&$1.5\times 10^{12}$& $5.1\times 10^{-7}$ & 0.1 &3.2\\ \hline
 $2.5\times 10^{23}$&$8.4\times 10^{14}$& $2.5\times 10^{-7}$ & 0.05 & 1.6\\ \hline
 $1.5\times 10^{30}$&$5.02\times 10^{21}$ & $5.1\times 10^{-8}$ & 0.01 &0.3\\ 
 \hline
\hline 
\multicolumn{5}{|c|}{Scenario B}\\
\hline
 $5.9\times 10^{-6}$&$2.0\times 10^{-14}$& $8.4\times 10^{-4}$ & 10
 & 650\\ \hline
 $1.3\times 10^{-2}$&$4.6\times 10^{-11}$& $4.2\times 10^{-4}$ & 5 & 312\\ \hline
 $10.1$&$3.4\times 10^{-8}$& $2.5\times 10^{-4}$ & 3 & 184\\ \hline
 $1.6\times 10^3$&$5.3\times 10^{-6}$& $1.7\times 10^{-4}$ & 2 & 122\\ \hline
 $9.8\times 10^{5}$&$3.2\times 10^{-3}$& $8.5\times 10^{-5}$ & 1 & 60.5\\ \hline
 $1.7\times 10^{9}$&$5.8$& $4.3\times 10^{-5}$ & 0.5 & 30\\ \hline
 $3.2\times 10^{15}$&$10.8\times 10^{6}$& $1.7\times 10^{-5}$ & 0.2 & 12\\ \hline
 $1.6\times 10^{18}$&$5.5\times 10^{9}$& $8.6\times 10^{-6}$ & 0.1 & 6\\ \hline
 $8.5\times 10^{20}$&$2.8\times 10^{12}$& $4.3\times 10^{-6}$ & 0.05 & 3.0\\ 
 \hline
 $5.4\times 10^{27}$&$1.8\times 10^{19}$ & $8.6\times 10^{-7}$ & 0.01 &0.6\\
 \hline
 \hline 
 \multicolumn{5}{|c|}{Scenario C}\\
 \hline
 $4.9\times 10^{-6}$&$1.6\times 10^{-14}$& $9.1\times 10^{-4}$ & 10
 & 820\\ \hline
 $1\times 10^{-2}$&$3.6\times 10^{-11}$& $4.6\times 10^{-4}$ & 5 & 397\\ \hline
 $7.9$&$2.6\times 10^{-8}$& $2.8\times 10^{-4}$ & 3 & 235\\ \hline
 $1.2\times 10^3$&$4.3\times 10^{-6}$& $1.9\times 10^{-4}$ & 2 & 156\\ \hline
 $7.8\times 10^{5}$&$2.6\times 10^{-3}$& $9.5\times 10^{-5}$ & 1 & 77.5\\ \hline
 $1.4\times 10^{9}$&$4.8$& $4.7\times 10^{-5}$ & 0.5 & 38.6\\ \hline
 $2.6\times 10^{15}$&$8.6\times 10^{6}$& $1.9\times 10^{-5}$ & 0.2 & 15.4\\ \hline
 $1.3\times 10^{18}$&$4.4\times 10^{9}$& $9.6\times 10^{-6}$ & 0.1 & 7.7\\ \hline
 $6.9\times 10^{20}$&$2.3\times 10^{12}$& $4.8\times 10^{-6}$ & 0.05 & 3.8\\ \hline
 $4.5\times 10^{27}$&$1.5\times 10^{19}$ & $9.6\times 10^{-7}$ & 0.01 &0.7\\ \hline
\end{tabular} 
\caption{Values of lifetime and coupling intensity ($g_{A_1H_1Z}=ig_3(P_A^\mu+P_1^\mu)$) for different values of $\mu_{12}^2$ or $(\Delta_h)$. The other parameters values are: ({Scenario A}) $m_{H_1}=32~{\rm GeV}$, $\Delta_n=10~{\rm GeV}$, $g_1\approx 10^{-4}$ and $g_2\approx 10^{-3}$; ({Scenario B}) $m_{H_1}=60~{\rm GeV}$, $\Delta_n=24~{\rm GeV}$, $g_1\approx 10^{-4}$ and $g_2\approx 10^{-2}$; ({Scenario C}) $m_{H_1}=77~{\rm GeV}$, $\Delta_n=7.5~{\rm GeV}$, $g_1\approx 10^{-4}$ and $g_2\approx 10^{-2}.$}
\label{results1}
\end{table}
\section{Numerical results}
 In accordance with the previous section, we can define three $\mu_{12}^2$ stages for the scenarios A, B and C, which are further discussed below in relation to their possible signatures at colliders, chiefly, a $e^+e^-$ one, like the
 International Linear Collider (ILC)\cite{ILCInternationalDevelopmentTeam:2022izu}.
 
\subsection{Two-component DM: $A_1$ stable or with a very long lifetime}

Since the Hermaphrodite DM states $ H_1 $ and $ A_1 $  could be protected from decaying into SM particles by the exact conservation of  $ Z_3 $ symmetry, with $\mu_{12}^2=0$ and $\Delta_h=0$, being the two particles stable in this case, hence,  DM candidates,  having the same mass values. In  Figure \ref{micro}, these cases are shown by red crosses. The ranges of masses are 25 GeV$\leq m_{H_1} \leq 35$ GeV, as well as among 60--67 GeV or 75--78 GeV.    
When the soft breaking term $\mu_{12}^2 $ is considered, the mass spectrum is no longer degenerate as 
$ m_ {H_1} <m_ {A_1} <m_ {A_2} <m_ {H_2} $.
The particles $ A_2 $, $ H_2 $ and $A_1$ are unstable and decay into $ H_1$ and SM particles. As mentioned in the previous section, the $A_1$ decay is induced at the one-loop level by the vertex $A_1 H_1 Z$. The decay width is controlled by the $\mu_{12}^2$ term. 
In  Figure \ref{time1}, taking values of $\mu_{12}^2$  between 0 and 10 GeV$^2$ (or $0\leq \Delta_h\leq 0.3$ GeV),  one can get a particle $A_1$ whose lifetime $\tau_{A_1}= 1/ \Gamma_{A_1}$ is around the age of the early universe when the stars began to form, after BBN.  In Figure \ref{time2} we can see that the relic density associated with $A_1$ is suppressed 
for larger values of $\Delta_H$ and $\mu_{12}^2$. As for collider studies, in this case, signatures are just like those in the I(2+1)HDM, the $Z_2\times Z'_2$ symmetry \cite{Hernandez-Sanchez:2022dnn,Hernandez-Sanchez:2020aop}.

\subsection{Decaying $A_1$ with displaced vertex inside a collider detector}

\begin{figure}
    \centering
    \includegraphics[width=8cm, height=6cm]{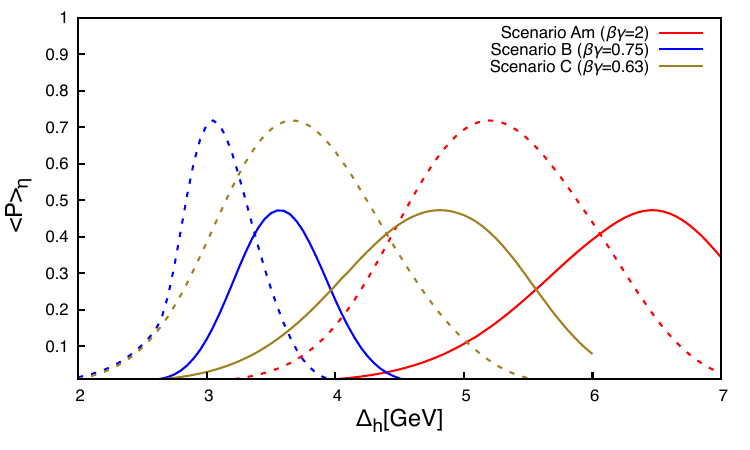}
    \includegraphics[width=8cm, height=6cm]{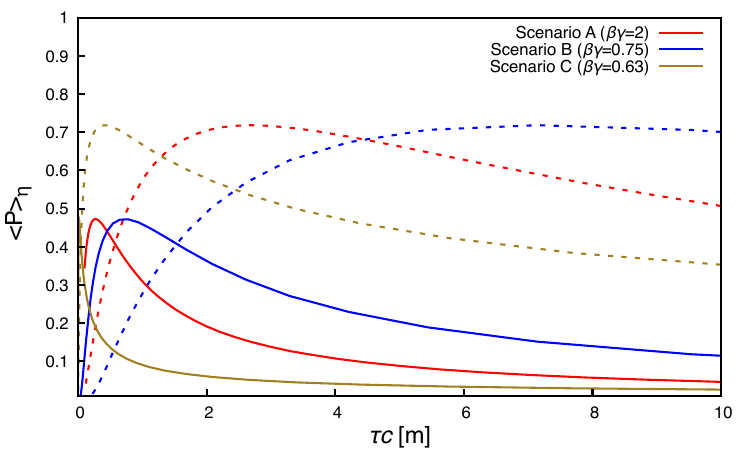}
    \caption{Average decay probability of the scalar $A_1$ for the three scenarios (A, B and C) discussed in the text. The continuous line is related to region 1 of  the detector while   the dashed line is related to region 2 of it (see Ref.~\cite{Accomando:2016rpc}). Scenario A: $M_{H_1}=32$ GeV, $M_{H_2}=85$ GeV, $g_1\approx 10^{-4}$, $g_2\approx 10^{-3}$  and $\Delta_h=5.2(6.5)$ GeV related with region 1 (region 2). 
Scenario B: $M_{H_1}=60$ GeV, $M_{H_2}=120$ GeV, $g_1\approx 10^{-4}$, $g_2\approx 10^{-2}$  and $\Delta_h=3(3.6)$ GeV for region 1(2). 
Scenario C: $M_{H_1}=77$ GeV, $M_{H_2}=84$ GeV, $g_1\approx 10^{-4}$, $g_2\approx 10^{-2}$  and $\Delta_h=3.7(4.8)$ GeV in the region 1(2). }
    \label{fig:enter-label}
\end{figure}

Now, we study the case when the particle $A_1$ is not cosmologically stable,  but it is stable enough to decay inside the detector with a displaced vertex. Specifically, we concentrate on the 
the possibility that such a state leaves traces in the detector due its decays to leptons ($l$'s, where $l=e,\mu$) or/and jets ($j$'s) plus  missing transverse energy ($\cancel{E}_T$). In fact, 
following Ref. \cite{Accomando:2016rpc}, we get the  average $A_1$ decay probability $\langle P\rangle_\eta$ as a function of  $\Delta_h$ and $\tau c$
in  Figure \ref{fig:enter-label},  wherein the dashed line (solid line) represents represent decays mostly inside  the detector (outside  the detector) while the red colour line refers to  Scenario A, the blue colour line refers to  Scenario B and  the brown colour line refers Scenario C.

\begin{figure}[tbh]
	\centering
\includegraphics[width=0.22\textwidth]{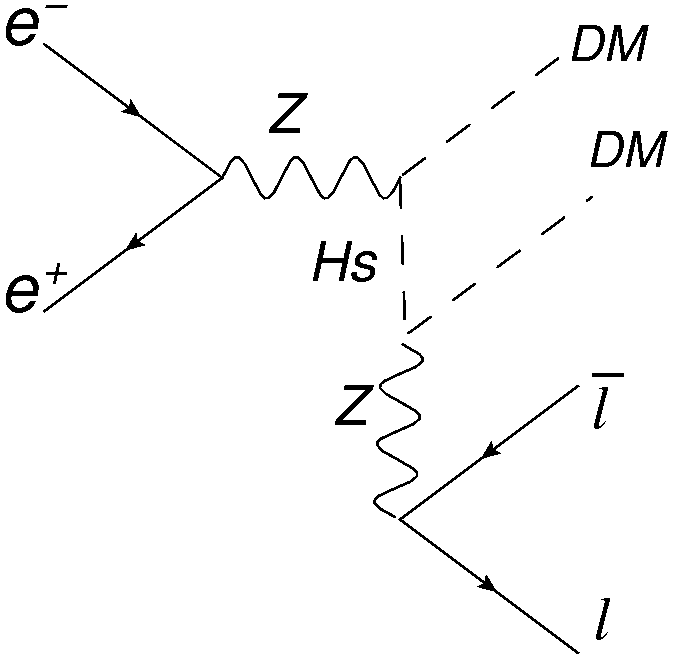} \hspace{.5cm}\vspace{.5cm}
	\includegraphics[width=0.21\textwidth]{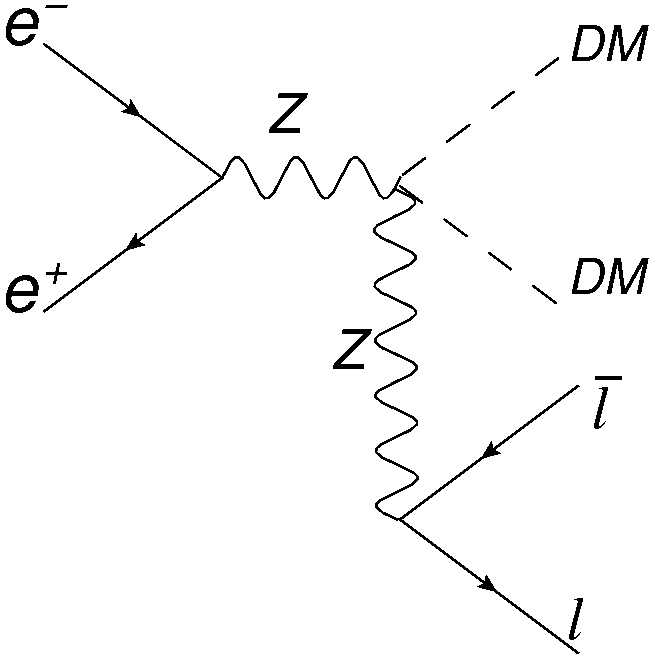}\hspace{.5cm}
	\includegraphics[width=0.3\textwidth]{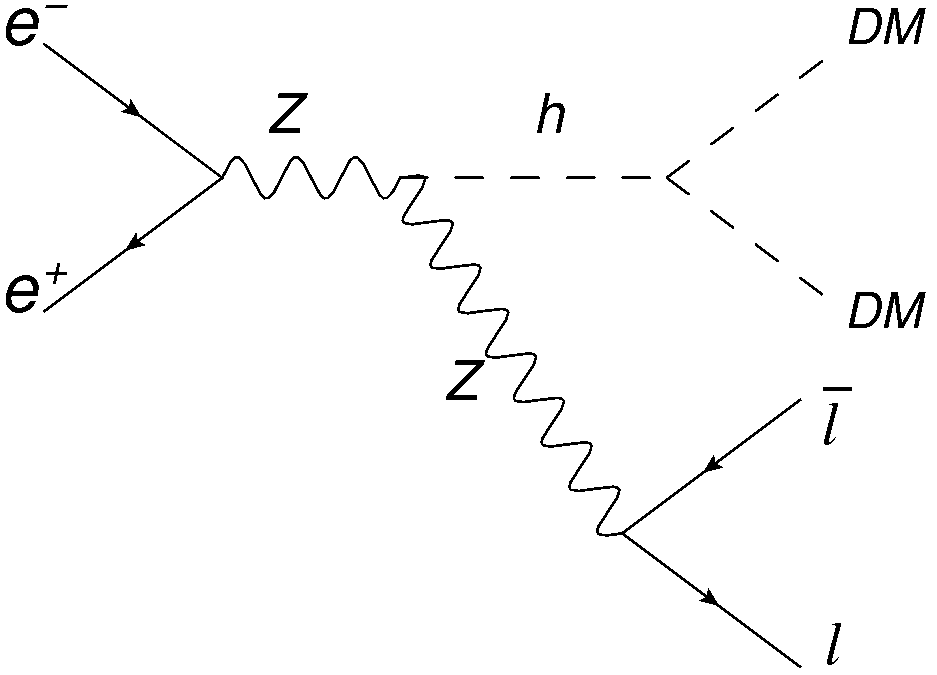} \hspace{1.5cm}
		\includegraphics[width=0.27\textwidth]{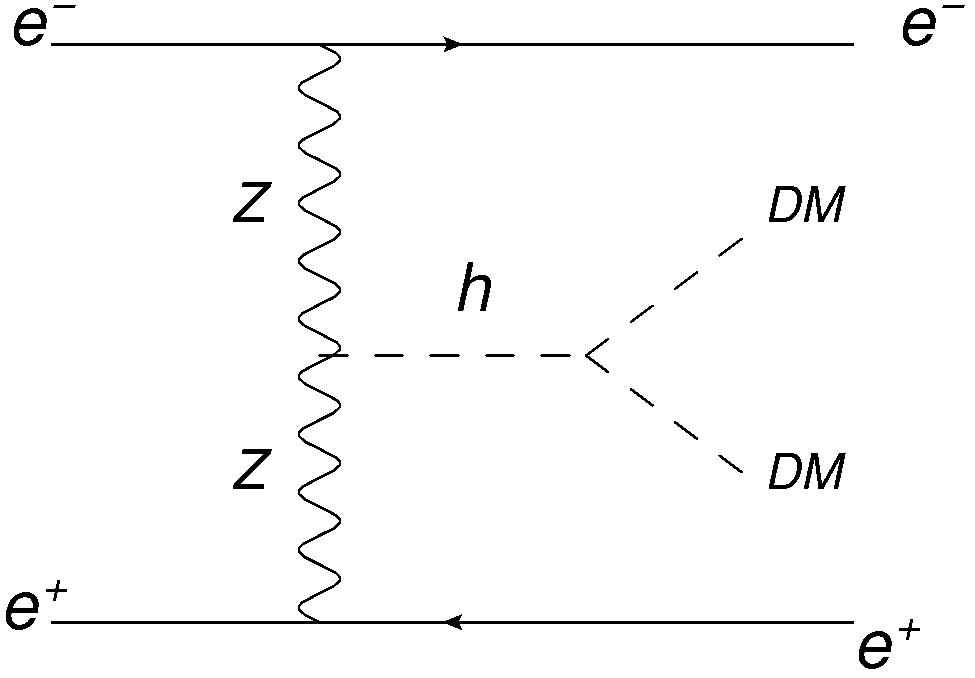} \hspace{.5cm}
	\includegraphics[width=0.21\textwidth]{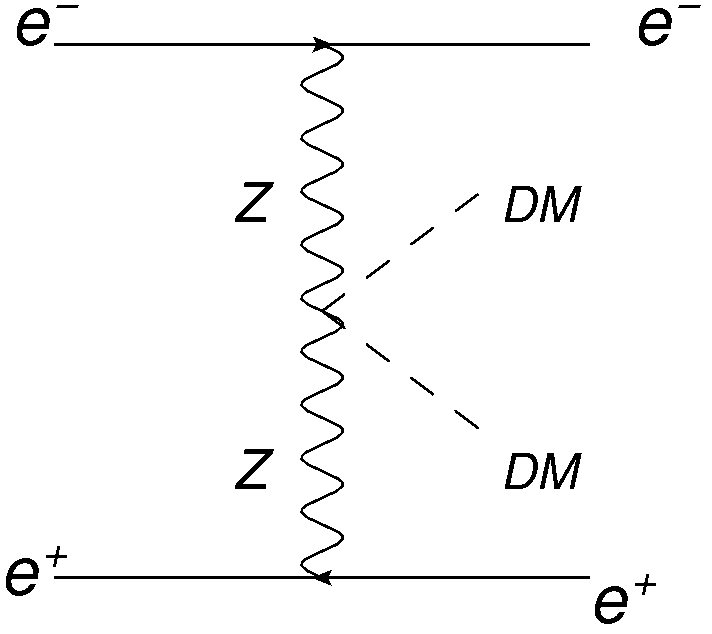}\hspace{.5cm} 
	\includegraphics[width=0.21\textwidth]{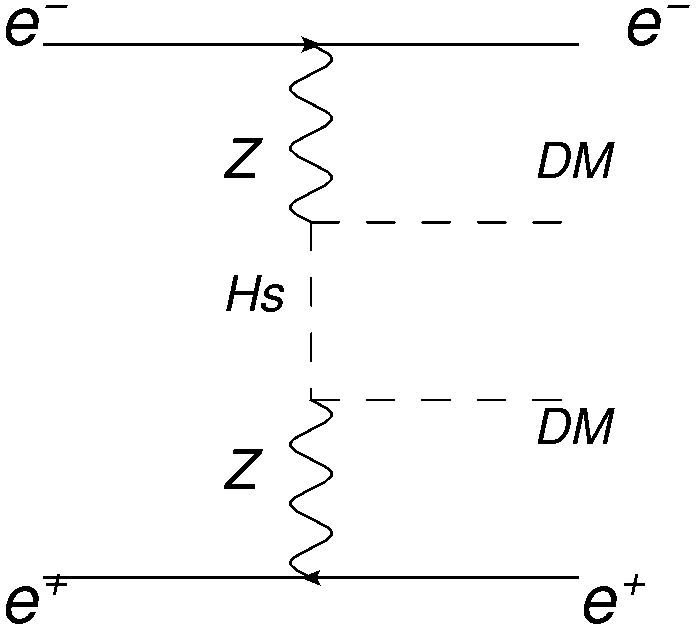} 
	\caption{Feynman diagrams for the processes $e^+e^-\to 2l+2~{\rm DM}$, where $l = e ^ {-} (\mu^-) $ and $ \bar {l} = e ^ {+} (\mu^+) $, $H_s=A_1,A_2$ and ${\rm DM}=H_1$. (The diagrams in the second row only enter for the case $l=e^-$ and $\bar l=e^+$). Here, we have also included irreducible background graphs, i.e., those not including $A_1$, however, they are negligible.}
    \label{eeto2l2h}
\end{figure}

\begin{figure}[tbh]
	\centering
	\includegraphics[width=14cm]{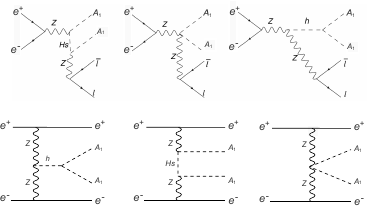} 
	\caption{Feynman diagrams for the processes $e^+ e^-\to 2l+2A_1$, where $l = e ^ {-} (\mu^-) $ and $ \bar {l} = e ^ {+} (\mu^+) $ and $H_s=H_1,H_2$, followed by $A_1A_1 \to 2l+2j + \cancel{E}_T$ or    $4l +\cancel{E}_T$. (The diagrams in the second row only enter for the case $l=e^-$ and $\bar l=e^+$). Here, we have not included any irreducible background graphs.   }
    \label{diagram-A_1}
\end{figure}

 As mentioned, we study DM production at the ILC, in particular, we focus on the processes $e^+e^-\to 2l + \cancel{E}_T$,  $4l + 2j+ \cancel{E}_T$ and  $6l + \cancel{E}_T$, with the latter two final states being very clean, thus offering spectacular signals of the model in the configuration considered \footnote{Note that, here, we neglect considering the case $e^+e^-\to 2l + 4j+ \cancel{E}_T$, as it will be polluted by significant QCD backgrounds.}.   (Hereafter, we use the shorthand notation $2l$ to signify an electron or muon pair of opposite charge and multiples thereof.)
The aforementioned processes are captured via Feynman diagrams as follows:  
for $ e ^ + e^- \to 2l + \cancel{E}_T$ see Figure \ref{eeto2l2h}, wherein the DM label refers to $H_1$, whereas for the processes  
$e^+e^-\to 4l + 2j+ \cancel{E}_T$ and  $6l + \cancel{E}_T$ see Figure \ref{diagram-A_1}, wherein the $A_1A_1$ pair decays to $2l+2j + \cancel{E}_T$ or    $4l +\cancel{E}_T$.

To study our processes, we have selected from  Table \ref{results1} one particular BP,  Scenario B with $\Delta_h=5$ GeV (recall that $\Delta_n = 24$ GeV), which will then feature an average displacement of $1.3 \times 10^{-2}$ m. As possible energy of the future 
$e^+e^-$ linear collider chosen we have adopted $\sqrt s_{ee}=1000$ GeV whereas the luminosity is   $L=1000$ fb$^{-1}$. 
We also assume the following beam polarisations:  80\% for the $e^-$ beam and 30\%  for the $e^+$ beam, although neither of these is necessary to uphold our forthcoming
conclusions.  MadGraph \cite{Alwall:2014hca} is used for our calculations at the parton level with integrated and differential distributions obtained via MadAnalysis  \cite{Conte:2012fm}.

\begin{figure}
  \begin{center}
    \includegraphics[scale=0.45]{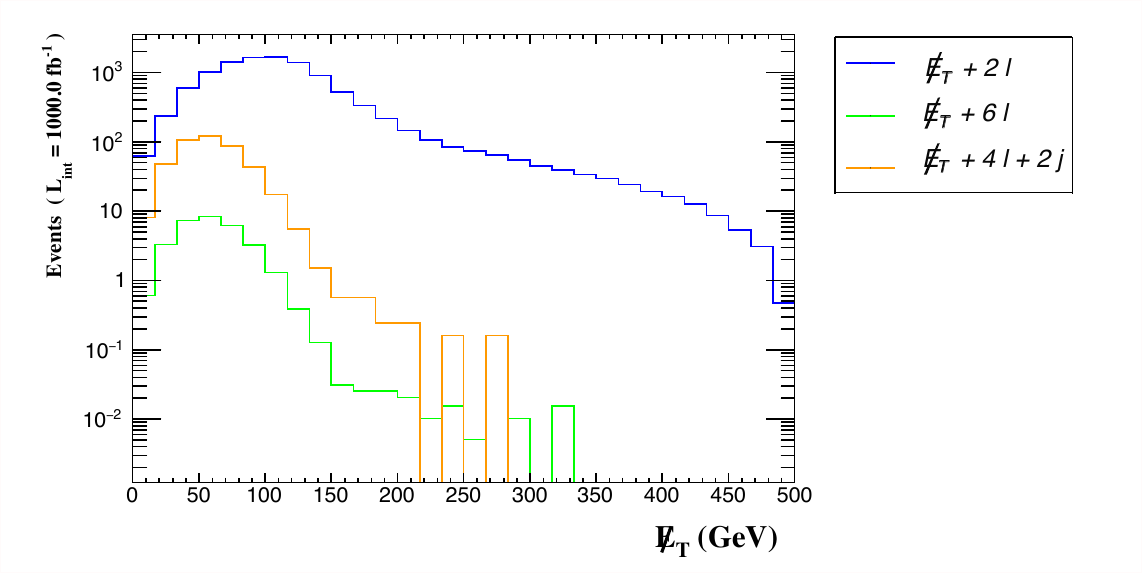}
     \includegraphics[scale=0.45]{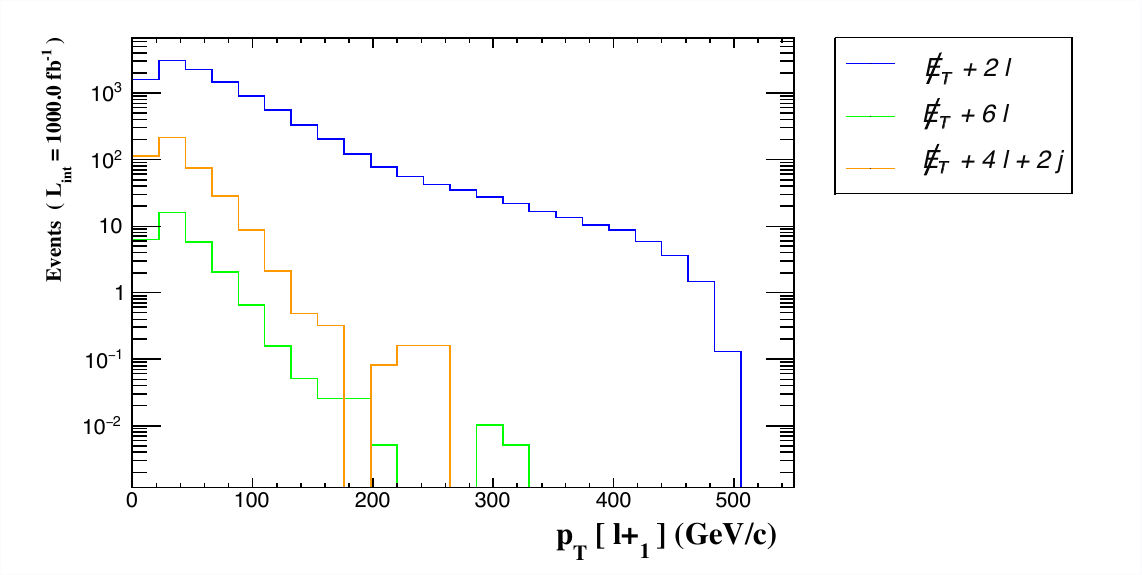}
\caption{Spectra in missing transverse energy (left) and transverse momentum of a lepton (right) for the processes discussed in the text and our chosen BP.
}
 \label{met-ILC-1}
  \end{center}
\end{figure}

\begin{figure}
  \begin{center}
    \includegraphics[scale=0.45]{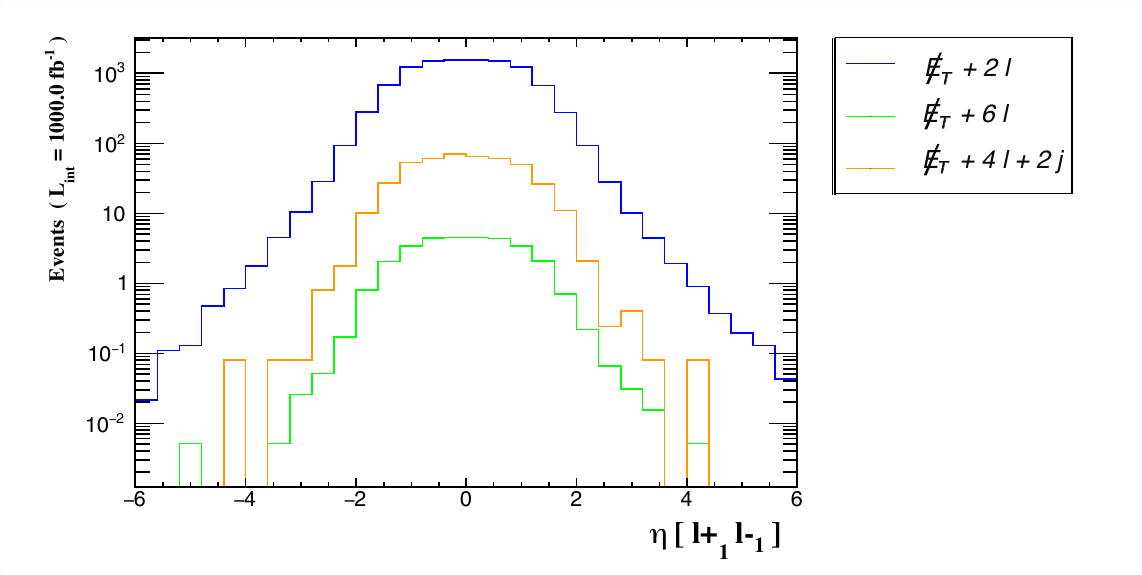}
     \includegraphics[scale=0.45]{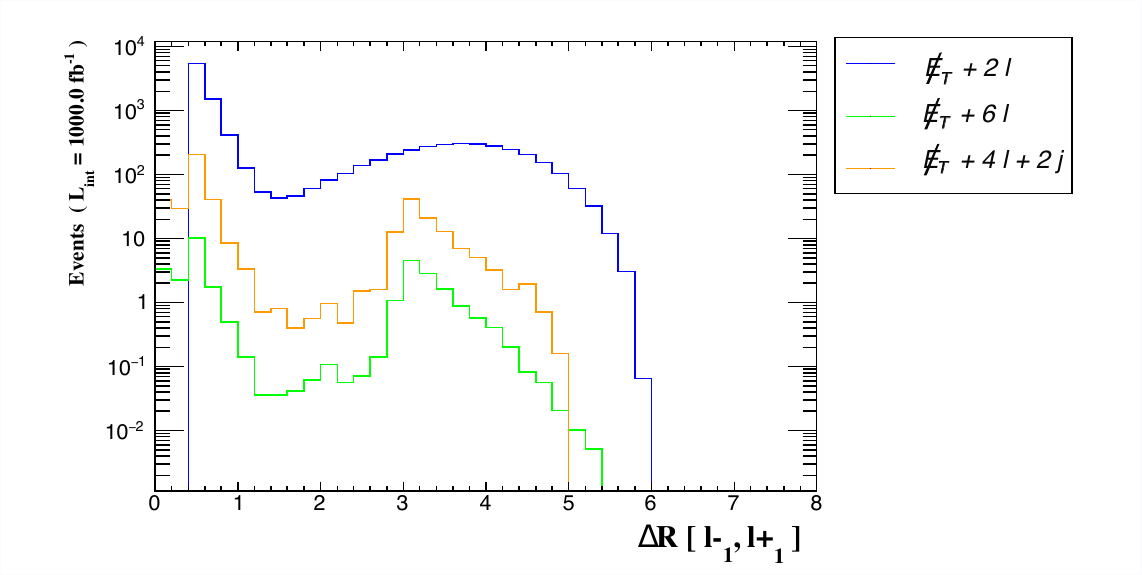}
 \caption{Spectra in pseudorapidity of a lepton (left) and separation between oppositely charged leptons of same flavour (right)  for the processes discussed in the text and our chosen BP.
}
  \label{deltar-ILC}
  \end{center}
\end{figure}

\begin{figure}
\begin{center}
    \includegraphics[scale=0.5]{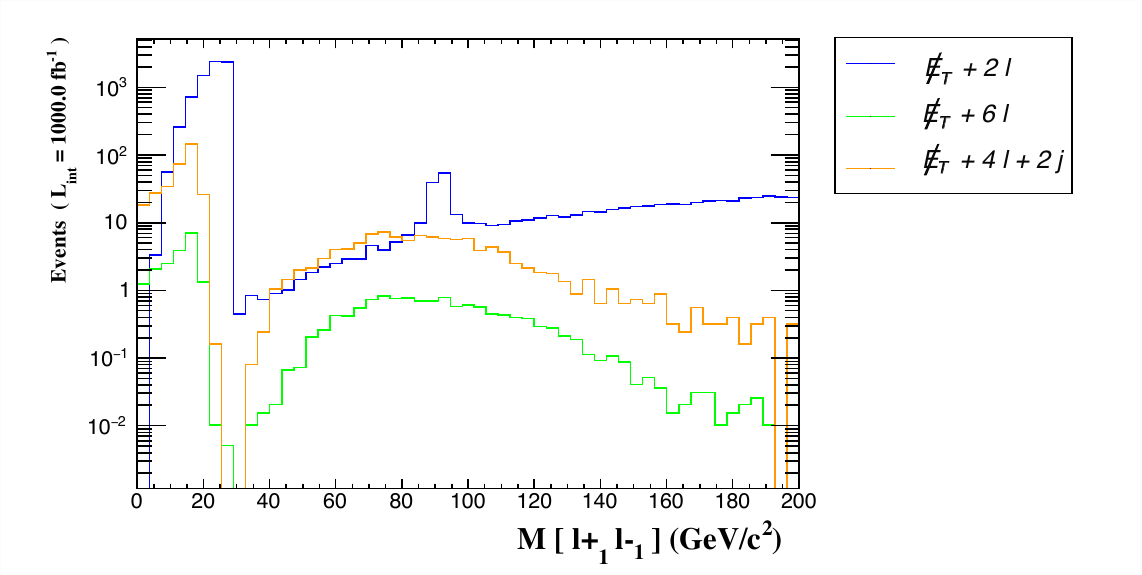}
\caption{Spectra in  invariant mass of pairs of oppositely charged leptons of same flavour  for the processes discussed in the text and our chosen BP.
}
  \label{mll-1-ILC}
  \end{center}
\end{figure}

\begin{figure}
\begin{center}
    \includegraphics[scale=0.5]{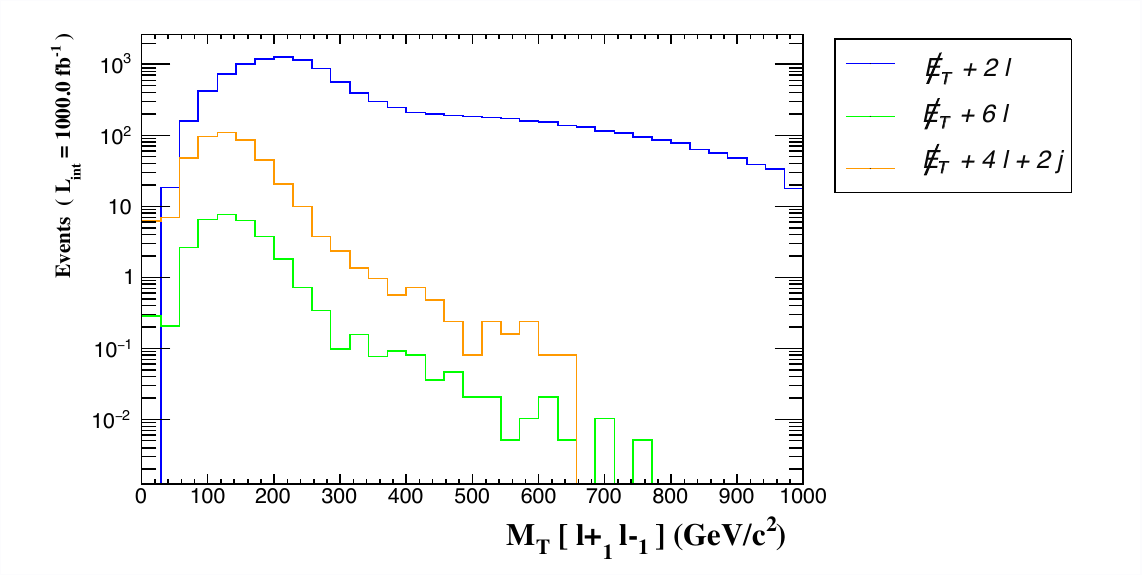}\\
    \includegraphics[scale=0.5]{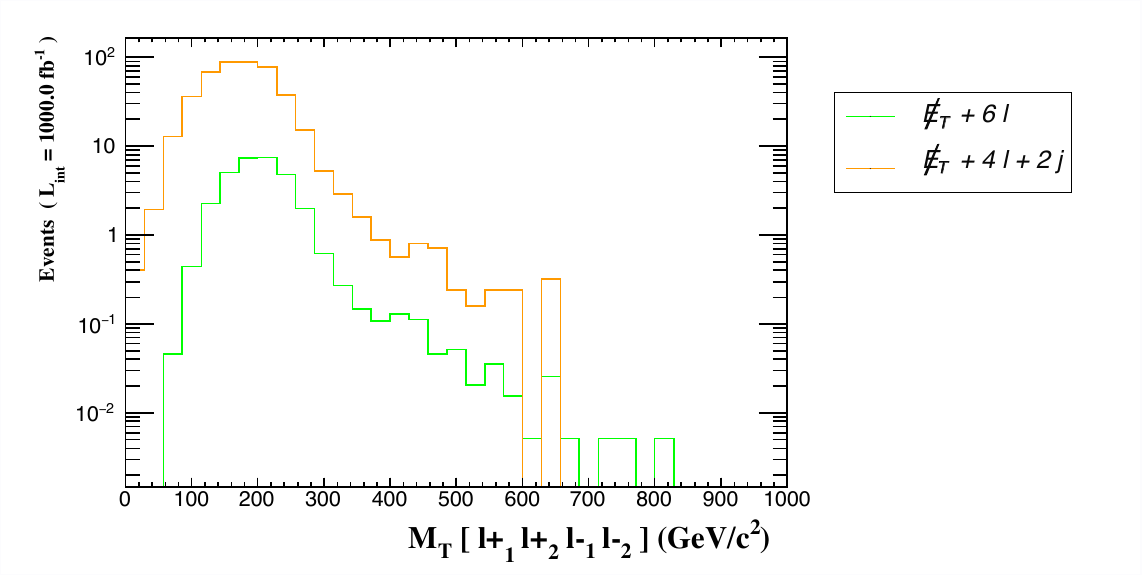}\\
    \includegraphics[scale=0.5]{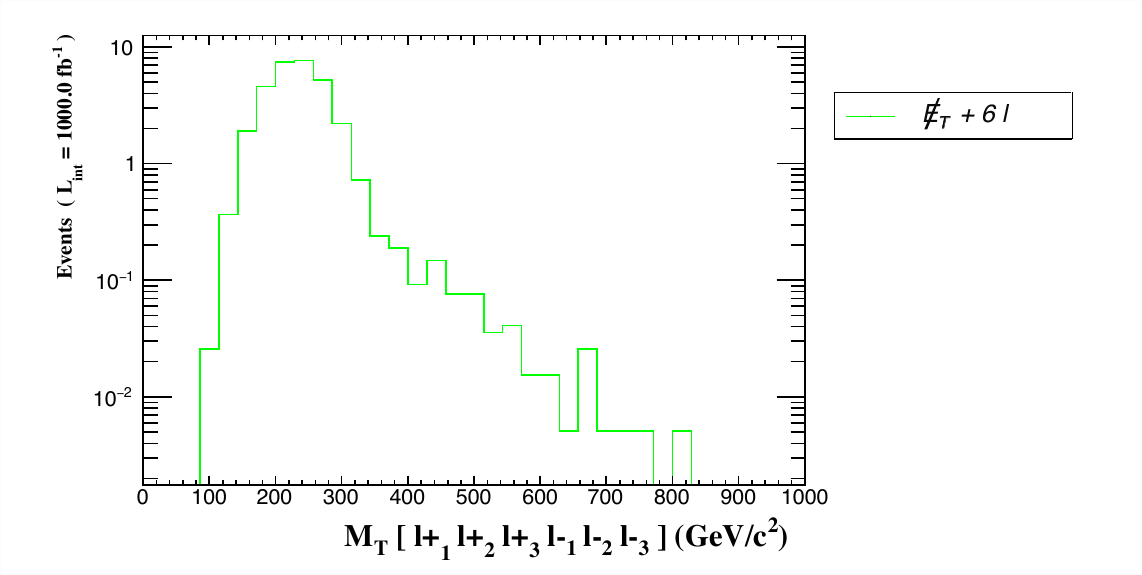}
\caption{Spectra in transverse mass of multiple pairs of oppositely charged leptons of same flavour  for the processes discussed in the text and our chosen BP.
}
  \label{mT-ILC}
  \end{center}
\end{figure}

 Table \ref{sigmaH_1A_1-ILC} shows the cross-sections of the various processes considered here. Here, the one for $ e^+ e^- \to 6 l +  \cancel{E}_T$ is calculated as $ \sigma(e^+ e^- \to 2l +  2 A_1) \times {\rm BR}(A_1 \to 2l + H_1)^2$ while the one for $e^+ e^- \to 4 l + 2j+ \cancel{E}_T$ as 
$ \sigma(e^+ e^- \to 2l +  2 A_1) \times {\rm BR}(A_1 \to 2l + H_1) \times {\rm BR}(A_1 \to 2j + H_1)$. 
Furthermore, we have investigated their differential appearance in terms of the distributions in Figures~\ref{met-ILC-1}--\ref{mT-ILC}. It is evident that signals of  Hermaphrodite DM are potentially accessible at the ILC and have rather distinctive kinematic features, which in turn depend on the values of the $H_1$ and $A_1$ as well as $A_2$ and $H_2$ masses, which could be used for characterisation purposes of potential excesses.
\begin{table}[htp]
\begin{center}
\begin{tabular}{|c|c|c|c|}
\hline
$X + \cancel{E}_T $ & $\sigma (e^+ e^- \to X + \cancel{E}_T) $ & Event rates \\ \hline
$2l+ 2H_1 $ & 10.8 fb & $1.08 \times 10^{4}$ \\ \hline
$4l + 2j+  2 H_1 $ & 0.442 fb & $4.42 \times 10^{2}$ \\ \hline
$6l +  2 H_1 $ & 0.031 fb & $31$ \\ \hline
\end{tabular}
\end{center}
	\caption{Cross-sections for the processes $e^+e^- \to 2 l + 2 H_1$, $e^+e^- \to 4 l + 2j+ 2 {H_1}$ and $e^+e^- \to 6 l + 2{H_1}$, where $2H_1$ would yield $\cancel{E}_T$,  for  $m_{H_1} = 60 $ GeV, $m_{A_1} = 65 $ GeV,  $m_{A_2} = 84 $ GeV, $m_{H_2} = 89 $ GeV, $g_1=0.0001$ and  other parameters as in Scenario B.}
	\label{sigmaH_1A_1-ILC}
\end{table}

\section{Conclusions}
In this paper,  we have studied a realisation of the 3HDM, wherein  one doublet is active, and two are inert (hence it is termed I(2+1)HDM), which, in the presence of a softly-broken $Z_3$ symmetry, yields two potential DM candidates, in the form of the lightest CP-even and CP-odd states from the inert sector, $H_1$ and $A_1$, respectively (wherein the label $1$ indicates that they emerge from the same inert doublet, although the actual doublet is irrelevant for phenomenological purposes). 

These two states, emerging from the inert sector (hence with the same $Z_3$ properties), while having opposite CP quantum numbers, are not mass degenerate and have different gauge couplings, so that they cannot be ascribed to being the real and imaginary part of a single complex field. Therefore, they have been aptly called Hermaphrodite DM in past literature. Such a mass difference emerges from the
soft-breaking of the $Z_3$ symmetry, and we have arranged the I(2+1)HDM parameters such that the  $A_1$ state is heavier than the $H_1$ (although the reverse is also possible, with no significant difference in the ensuing phenomenology).

If the breaking scale is small enough, this model provides a long-lived $A_1$ state, such that it could be another `effective' DM candidate (i.e., have a lifetime comparable to the age of the universe).  In this case, we have delineated the regions of parameter space of our BSM construct where the 
relic density can be saturated by either $H_1$ or both $H_1$ and $A_1$, depending on the lifetime of $A_1$ and the universe history, while accounting for current constraints from both DD and ID searches for DM signals. The key result is that our BSM scenario offers a variety of parameter space configurations 
satisfying either condition, while also potentially affecting cosmological dynamics, although we have not studied the latter aspect in detail here. 

In contrast, when the $A_1$ lifetime is short, such a state could decay into the DM particle and (via a $Z$ boson) additional SM objects inside a typical collider detector, thereby
yielding interesting signatures with displaced tracks accompanying the 
usual $\cancel{E}_T$. In such a case, we have looked at the possibilities offered by the ILC, including exploiting its polarisation capabilities, of accessing 
signals involving multi-leptons in the final state, confirming that significant sensitivity exists for their detection. Furthermore, we have highlighted the distinctive shape that a variety of observables would have in charactering these
signatures, thus offering the ability to eventually pinpoint this BSM scenario. We have proven this to be the case through a parton-level analysis, albeit without a signal-to-background study, that we leave for a future endeavour.

\section*{Acknowledgements}
SM acknowledges support from the STFC Consolidated Grant ST/X000583/1 and is partially financed through the NExT Institute. TS is supported in part by the JSPS KAKENHI Grant Number 20H00160. TS and SM are partially supported by the Kogakuin University Grant for the  research project ``Phenomenological study of new physics models with extended Higgs sector''. JH-S and CG-H acknowledge the support by SNII-SECIHTI, VIEP- BUAP and PRODEP-SEP, SECIHTI -CBF-2025-G-1187. as well as, JHS  acknowledges SECIHTI Sabbatical Support 2025 under the grant ``Higgs and Dark Matter Physics''. 

\bibliographystyle{apsrev4-2.bst}
\bibliography{Z3-BIB}

@article{ILCInternationalDevelopmentTeam:2022izu,
    author = "Aryshev, Alexander and others",
    collaboration = "ILC International Development Team",
    title = "{The International Linear Collider: Report to Snowmass 2021}",
    eprint = "2203.07622",
    archivePrefix = "arXiv",
    primaryClass = "physics.acc-ph",
    reportNumber = "DESY-22-045, IFT-UAM/CSIC-22-028, KEK Preprint 2021-61, IFT--UAM/CSIC--22-028, KEK Preprint 2021-61,
  PNNL-SA-160884, SLAC-PUB-17662, FERMILAB-FN-1171-PPD-QIS-SCD-TD, PNNL-SA-160884",
    doi = "10.2172/1873702",
    journal = "",
    month = "3",
    year = "2022"
}

@article{ATLAS:2012yve,
    author = "Aad, Georges and others",
    collaboration = "ATLAS",
    title = "{Observation of a new particle in the search for the Standard Model Higgs boson with the ATLAS detector at the LHC}",
    eprint = "1207.7214",
    archivePrefix = "arXiv",
    primaryClass = "hep-ex",
    reportNumber = "CERN-PH-EP-2012-218",
    doi = "10.1016/j.physletb.2012.08.020",
    journal = "Phys. Lett. B",
    volume = "716",
    pages = "1--29",
    year = "2012"
}

@article{CMS:2012qbp,
    author = "Chatrchyan, Serguei and others",
    collaboration = "CMS",
    title = "{Observation of a New Boson at a Mass of 125 GeV with the CMS Experiment at the LHC}",
    eprint = "1207.7235",
    archivePrefix = "arXiv",
    primaryClass = "hep-ex",
    reportNumber = "CMS-HIG-12-028, CERN-PH-EP-2012-220",
    doi = "10.1016/j.physletb.2012.08.021",
    journal = "Phys. Lett. B",
    volume = "716",
    pages = "30--61",
    year = "2012"
}

@article{Barger:1989fj,
    author = "Barger, Vernon D. and Hewett, J. L. and Phillips, R. J. N.",
    title = "{New Constraints on the Charged Higgs Sector in Two Higgs Doublet Models}",
    reportNumber = "MAD-PH-530",
    doi = "10.1103/PhysRevD.41.3421",
    journal = "Phys. Rev. D",
    volume = "41",
    pages = "3421--3441",
    year = "1990"
}

@article{Grossman:1994jb,
    author = "Grossman, Yuval",
    title = "{Phenomenology of models with more than two Higgs doublets}",
    eprint = "hep-ph/9401311",
    archivePrefix = "arXiv",
    reportNumber = "WIS-94-3-PH",
    doi = "10.1016/0550-3213(94)90316-6",
    journal = "Nucl. Phys. B",
    volume = "426",
    pages = "355--384",
    year = "1994"
}

@article{Aoki:2009ha,
    author = "Aoki, Mayumi and Kanemura, Shinya and Tsumura, Koji and Yagyu, Kei",
    title = "{Models of Yukawa interaction in the two Higgs doublet model, and their collider phenomenology}",
    eprint = "0902.4665",
    archivePrefix = "arXiv",
    primaryClass = "hep-ph",
    reportNumber = "TU-839, UT-HET-022, IC-2009-007",
    doi = "10.1103/PhysRevD.80.015017",
    journal = "Phys. Rev. D",
    volume = "80",
    pages = "015017",
    year = "2009"
}

@article{Deshpande:1977rw,
    author = "Deshpande, Nilendra G. and Ma, Ernest",
    title = "{Pattern of Symmetry Breaking with Two Higgs Doublets}",
    reportNumber = "OITS-81",
    doi = "10.1103/PhysRevD.18.2574",
    journal = "Phys. Rev. D",
    volume = "18",
    pages = "2574",
    year = "1978"
}

@article{Aranda:2019vda,
    author = "Aranda, A. and Hern{\'a}ndez-Otero, D. and Hern{\'a}ndez-Sanchez, J. and Keus, V. and Moretti, S. and Rojas-Ciofalo, D. and Shindou, T.",
    title = "{Z$_3$ symmetric inert ( 2+1 )-Higgs-doublet model}",
    eprint = "1907.12470",
    archivePrefix = "arXiv",
    primaryClass = "hep-ph",
    doi = "10.1103/PhysRevD.103.015023",
    journal = "Phys. Rev. D",
    volume = "103",
    number = "1",
    pages = "015023",
    year = "2021"
}

@article{Ivanov:2012,
doi = {10.1088/1751-8113/45/21/215201},
url = {https://doi.org/10.1088/1751-8113/45/21/215201},
year = {2012},
month = {may},
publisher = {IOP Publishing},
volume = {45},
number = {21},
pages = {215201},
author = {Ivanov, Igor P and Keus, Venus and Vdovin, Evgeny},
title = {Abelian symmetries in multi-Higgs-doublet models},
journal = {Journal of Physics A: Mathematical and Theoretical}
}

@article{Keus:2014isa,
    author = "Keus, Venus and King, Stephen F. and Moretti, Stefano",
    title = "{Phenomenology of the inert ( 2+1 ) and ( 4+2 ) Higgs doublet models}",
    eprint = "1408.0796",
    archivePrefix = "arXiv",
    primaryClass = "hep-ph",
    doi = "10.1103/PhysRevD.90.075015",
    journal = "Phys. Rev. D",
    volume = "90",
    number = "7",
    pages = "075015",
    year = "2014"
}

@article{Cao:2007rm,
    author = "Cao, Qing-Hong and Ma, Ernest and Rajasekaran, G.",
    title = "{Observing the Dark Scalar Doublet and its Impact on the Standard-Model Higgs Boson at Colliders}",
    eprint = "0708.2939",
    archivePrefix = "arXiv",
    primaryClass = "hep-ph",
    reportNumber = "UCRHEP-T437",
    doi = "10.1103/PhysRevD.76.095011",
    journal = "Phys. Rev. D",
    volume = "76",
    pages = "095011",
    year = "2007"
}

@article{Lundstrom:2008ai,
    author = "Lundstrom, Erik and Gustafsson, Michael and Edsjo, Joakim",
    title = "{The Inert Doublet Model and LEP II Limits}",
    eprint = "0810.3924",
    archivePrefix = "arXiv",
    primaryClass = "hep-ph",
    doi = "10.1103/PhysRevD.79.035013",
    journal = "Phys. Rev. D",
    volume = "79",
    pages = "035013",
    year = "2009"
}

@article{Aaboud:2019rtt,
    author = "Aaboud, Morad and others",
    collaboration = "ATLAS",
    title = "{Combination of searches for invisible Higgs boson decays with the ATLAS experiment}",
    eprint = "1904.05105",
    archivePrefix = "arXiv",
    primaryClass = "hep-ex",
    reportNumber = "CERN-EP-2019-046",
    doi = "10.1103/PhysRevLett.122.231801",
    journal = "Phys. Rev. Lett.",
    volume = "122",
    number = "23",
    pages = "231801",
    year = "2019"
}

@article{Cordero-Cid:2018man,
    author = "Cordero-Cid, A. and Hern{\'a}ndez-S{\'a}nchez, J. and Keus, V. and Moretti, S. and Rojas, D. and Soko{\l}owska, D.",
    title = "{Lepton collider indirect signatures of dark CP-violation}",
    eprint = "1812.00820",
    archivePrefix = "arXiv",
    primaryClass = "hep-ph",
    reportNumber = "DMIIP-2018",
    doi = "10.1140/epjc/s10052-020-7689-0",
    journal = "Eur. Phys. J. C",
    volume = "80",
    number = "2",
    pages = "135",
    year = "2020"
}

@article{Cordero:2017owj,
    author = "Cordero, A. and Hernandez-Sanchez, J. and Keus, V. and King, S. F. and Moretti, S. and Rojas, D. and Sokolowska, D.",
    title = "{Dark Matter Signals at the LHC from a 3HDM}",
    eprint = "1712.09598",
    archivePrefix = "arXiv",
    primaryClass = "hep-ph",
    doi = "10.1007/JHEP05(2018)030",
    journal = "JHEP",
    volume = "05",
    pages = "030",
    year = "2018"
}

@article{Mertig:1990an,
title = {Feyn Calc - Computer-algebraic calculation of Feynman amplitudes},
journal = {Computer Physics Communications},
volume = {64},
number = {3},
pages = {345-359},
year = {1991},
issn = {0010-4655},
doi = {https://doi.org/10.1016/0010-4655(91)90130-D},
url = {https://www.sciencedirect.com/science/article/pii/001046559190130D},
author = {R. Mertig and M. Böhm and A. Denner}
}

@article{Hahn:1998yk,
    author = "Hahn, T. and Perez-Victoria, M.",
    title = "{Automatized one loop calculations in four-dimensions and D-dimensions}",
    eprint = "hep-ph/9807565",
    archivePrefix = "arXiv",
    reportNumber = "UG-FT-87-98, KA-TP-7-1998",
    doi = "10.1016/S0010-4655(98)00173-8",
    journal = "Comput. Phys. Commun.",
    volume = "118",
    pages = "153--165",
    year = "1999"
}

@article{Passarino:1978jh,
    author = "Passarino, G. and Veltman, M. J. G.",
    title = "{One Loop Corrections for e+ e- Annihilation Into mu+ mu- in the Weinberg Model}",
    reportNumber = "Print-79-0284 (UTRECHT)",
    doi = "10.1016/0550-3213(79)90234-7",
    journal = "Nucl. Phys. B",
    volume = "160",
    pages = "151--207",
    year = "1979"
}

@article{plank2020,
    author = "Aghanim, N. and others",
    collaboration = "Planck",
    title = "{Planck 2018 results. VI. Cosmological parameters}",
    eprint = "1807.06209",
    archivePrefix = "arXiv",
    primaryClass = "astro-ph.CO",
    doi = "10.1051/0004-6361/201833910",
    journal = "Astron. Astrophys.",
    volume = "641",
    pages = "A6",
    year = "2020",
    note = "[Erratum: Astron.Astrophys. 652, C4 (2021)]"
}

@article{Aprile:2018dbl,
    author = "Aprile, E. and others",
    collaboration = "XENON",
    title = "{Dark Matter Search Results from a One Ton-Year Exposure of XENON1T}",
    eprint = "1805.12562",
    archivePrefix = "arXiv",
    primaryClass = "astro-ph.CO",
    doi = "10.1103/PhysRevLett.121.111302",
    journal = "Phys. Rev. Lett.",
    volume = "121",
    number = "11",
    pages = "111302",
    year = "2018"
}

@article{Karwin:2016tsw,
    author = "Karwin, Christopher and Murgia, Simona and Tait, Tim M. P. and Porter, Troy A. and Tanedo, Philip",
    title = "{Dark Matter Interpretation of the Fermi-LAT Observation Toward the Galactic Center}",
    eprint = "1612.05687",
    archivePrefix = "arXiv",
    primaryClass = "hep-ph",
    reportNumber = "UCI-HEP-TR-2016-22",
    doi = "10.1103/PhysRevD.95.103005",
    journal = "Phys. Rev. D",
    volume = "95",
    number = "10",
    pages = "103005",
    year = "2017"
}

@article{Belanger2015,
    author = "B{\'e}langer, G. and Boudjema, F. and Pukhov, A. and Semenov, A.",
    title = "{micrOMEGAs4.1: two dark matter candidates}",
    eprint = "1407.6129",
    archivePrefix = "arXiv",
    primaryClass = "hep-ph",
    doi = "10.1016/j.cpc.2015.03.003",
    journal = "Comput. Phys. Commun.",
    volume = "192",
    pages = "322--329",
    year = "2015"
}

@article{Alwall:2014hca,
    author = "Alwall, J. and Frederix, R. and Frixione, S. and Hirschi, V. and Maltoni, F. and Mattelaer, O. and Shao, H. -S. and Stelzer, T. and Torrielli, P. and Zaro, M.",
    title = "{The automated computation of tree-level and next-to-leading order differential cross sections, and their matching to parton shower simulations}",
    eprint = "1405.0301",
    archivePrefix = "arXiv",
    primaryClass = "hep-ph",
    reportNumber = "CERN-PH-TH-2014-064, CP3-14-18, LPN14-066, MCNET-14-09, ZU-TH-14-14",
    doi = "10.1007/JHEP07(2014)079",
    journal = "JHEP",
    volume = "07",
    pages = "079",
    year = "2014"
}

@article{Conte:2012fm,
    author = "Conte, Eric and Fuks, Benjamin and Serret, Guillaume",
    title = "{MadAnalysis 5, A User-Friendly Framework for Collider Phenomenology}",
    eprint = "1206.1599",
    archivePrefix = "arXiv",
    primaryClass = "hep-ph",
    reportNumber = "IPHC-PHENO-06",
    doi = "10.1016/j.cpc.2012.09.009",
    journal = "Comput. Phys. Commun.",
    volume = "184",
    pages = "222--256",
    year = "2013"
}

@article{LZ:2022lsv,
    author = "Aalbers, J. and others",
    collaboration = "LZ",
    title = "{First Dark Matter Search Results from the LUX-ZEPLIN (LZ) Experiment}",
    eprint = "2207.03764",
    archivePrefix = "arXiv",
    primaryClass = "hep-ex",
    doi = "10.1103/PhysRevLett.131.041002",
    journal = "Phys. Rev. Lett.",
    volume = "131",
    number = "4",
    pages = "041002",
    year = "2023"
}

@article{Belyaev:2012qa,
    author = "Belyaev, Alexander and Christensen, Neil D. and Pukhov, Alexander",
    title = "{CalcHEP 3.4 for collider physics within and beyond the Standard Model}",
    eprint = "1207.6082",
    archivePrefix = "arXiv",
    primaryClass = "hep-ph",
    reportNumber = "PITT-PACC-1209",
    doi = "10.1016/j.cpc.2013.01.014",
    journal = "Comput. Phys. Commun.",
    volume = "184",
    pages = "1729--1769",
    year = "2013"
}

@article{Accomando:2016rpc,
    author = "Accomando, Elena and Delle Rose, Luigi and Moretti, Stefano and Olaiya, Emmanuel and Shepherd-Themistocleous, Claire H.",
    title = "{Novel SM-like Higgs decay into displaced heavy neutrino pairs in U(1)' models}",
    eprint = "1612.05977",
    archivePrefix = "arXiv",
    primaryClass = "hep-ph",
    doi = "10.1007/JHEP04(2017)081",
    journal = "JHEP",
    volume = "04",
    pages = "081",
    year = "2017"
}

@article{Hernandez-Sanchez:2022dnn,
    author = "Hernandez-Sanchez, Jaime and Keus, Venus and Moretti, Stefano and Sokolowska, Dorota",
    title = "{Complementary collider and astrophysical probes of multi-component Dark Matter}",
    eprint = "2202.10514",
    archivePrefix = "arXiv",
    primaryClass = "hep-ph",
    reportNumber = "DIAS-STP-22-01",
    doi = "10.1007/JHEP03(2023)045",
    journal = "JHEP",
    volume = "03",
    pages = "045",
    year = "2023"
}

@article{Hernandez-Sanchez:2020aop,
    author = "Hernandez-Sanchez, J. and Keus, V. and Moretti, S. and Rojas-Ciofalo, D. and Sokolowska, D.",
    title = "{Complementary Probes of Two-component Dark Matter}",
    eprint = "2012.11621",
    archivePrefix = "arXiv",
    primaryClass = "hep-ph",
    reportNumber = "HIP-2020-35/TH, IIPDM-20",
    journal = "",
    month = "12",
    year = "2020"
}

\end{document}